\documentclass{article}
\usepackage[utf8]{inputenc}
\usepackage{physics}
\usepackage{bbold}
\usepackage{microtype,amssymb,amsmath,amsthm,mathrsfs,enumitem,hyperref,braket,bbm,graphicx,xcolor}
\usepackage{vmargin}
\usepackage{latexsym}
\setmarginsrb{2.5cm}{1cm}{2.5cm}{2cm}{12pt}{11mm}{0pt}{13mm}

\newcommand{\ma}[1]{{\color{red}{#1}}}

\title{Toward matter dynamics in spin foam quantum gravity}

\author{Masooma Ali${}^{a}$ \; and \; Sebastian Steinhaus${}^{b,}$\footnote{sebastian.steinhaus@uni-jena.de} \vspace{0.2cm} \\
${}^a$ Perimeter Institute For Theoretical Physics, 31 Caroline St N, Waterloo, ON N2L 2Y5 \\
${}^b$ Theoretisch-Physikalisches Institut, Friedrich-Schiller-Universit\"at Jena,\\ Max-Wien-Platz 1, 07743 Jena, Germany}

\begin{document}

\maketitle

\begin{abstract}{
Any approach to pure quantum gravity must eventually face the question of coupling quantum matter to the theory.
In the past, several ways of coupling matter to spin foam quantum gravity have been proposed, but the dynamics of the coupled matter-gravity system is challenging to explore.
To take first steps towards uncovering the influence quantum matter has on spin foam models, we couple free, massive scalar lattice field theory to a restricted, semi-classical 4d spin foam model, called quantum cuboids. This model can be understood as a superposition of hypercuboidal (and thus irregular) lattices. Both theories are coupled by defining scalar lattice field theory on irregular lattices via discrete exterior calculus and then superimposing these theories by summing over spin foam configurations. We compute expectation values of geometric and matter observables using Markov Chain Monte Carlo techniques. From the observables, we identify a regime in parameter space, in which the spin foam possesses a finite total volume and looks on average like a regular lattice with an emergent lattice spacing dependent on the mass of the scalar field. We also measure the 2-point correlation function and correlation length of the scalar field in relation to the geodesic distance encoded in the spin foam. Our results are consistent with the correlation function of ordinary scalar lattice field theory defined on a fixed regular lattice with the emergent lattice spacing and the same mass. We conclude that in this regime of the model, the scalar field is not sensitive to the fluctuations of the spin foam and effectively behaves as if it is defined on a fixed regular lattice.
}
\end{abstract}

\tableofcontents

\section{Introduction}

The definition of a consistent framework of gravity and quantum matter is one of the strongest reasons to search for a theory of quantum gravity. Indeed, at the classical level the dynamics of gravity and matter are deeply intertwined and it is expected that the same holds in the quantum regime of both theories. Therefore, all approaches to quantum gravity must eventually face the question how to include quantum matter in their approach, in particular those that describe pure quantum gravity. The purpose of this article is to make progress in this direction in spin foam quantum gravity and present for the first time expectation values of geometric and matter observables in a restricted setting.

Spin foam quantum gravity \cite{cbook,alexreview} is a background independent approach to quantum gravity, frequently referred to as the covariant / path integral formulation of loop quantum gravity \cite{thomasbook}. The starting point is Plebanski's formulation of general relativity \cite{plebanski}, as a constrained topological quantum field theory, where we first discretize topological BF theory on a combinatorial 2-complex \cite{bf-theory} and then impose the constraints. This theory decorates the discretization with group-theoretic, algebraic data encoding quantum geometric quantities, e.g. areas of triangles and (fuzzy) shapes of tetrahedra. In the path integral one then sums over these data to describe a sum over geometries. One of the most widely used and explored modern spin foam models for 4d gravity is the Engle-Pereira-Rovelli-Livine / Freidel-Krasnov (EPRL-FK) model \cite{Engle:2007qf,eprl,FK}, which has the particular advantage of connecting on its boundary to the kinematical loop quantum gravity Hilbert space.

Different scenarios on how to include matter have been explored in loop quantum gravity and spin foams, e.g. unification scenarios in which a large symmetry group encodes both gravity and matter \cite{Smolin:2007rx} or scenarios in which matter degrees of freedom are used to deparameterize the system \cite{Brown:1994py,Domagala:2010bm, Husain:2011tk,Han:2020chr}. In many cases, also in this work, one considers adding matter ``on top'' of the pure quantum gravity theory, here the spin foam. The idea is to use the geometry encoded in a spin foam as the (fluctuating) space-time on which matter degrees of freedom and their interactions are defined. These works include gauge fields \cite{oriti-pfeiffer,Mikovic:2001xi,Mikovic:2002uq,simone-3dYM} and fermions \cite{Bianchi:2010bn,Han:2011as}. Moreover, a spin foam model coupled to a massless scalar field exists \cite{Kisielowski:2018oiv}, which was derived from loop quantum gravity coupled to a scalar field following similar works in loop quantum cosmology \cite{Ashtekar:2010ve,Henderson:2010qd}. Defining such coupled systems is vital to investigate how quantum matter and gravity mutually affect each other, and has given rise to interesting questions in other approaches to quantum gravity. One example is the question whether there are (unexpected) constraints from quantum gravity on the possible matter content of the universe or whether quantum gravity effects significantly influence the matter theory, e.g. the mass of elementary particles. Both effects have e.g. been seen in asymptotic safety \cite{asymptotic-safety1}, see e.g. \cite{Dona:2013qba,Eichhorn:2019tcj,Eichhorn:2022jqj}. In discrete approaches, studying such systems is challenging but possible, e.g. recently in Causal Dynamical Triangulations \cite{Ambjorn:2012jv} a scalar field was coupled to the gravitational theory and the authors observed a matter driven change of space-time topology \cite{Ambjorn:2021fkp,Ambjorn:2021uge}.
In spin foams the dynamics of such coupled systems, in particular in the deep quantum geometric regime, is barely known, yet more humble goals might be within reach. Indeed, such matter-gravity systems should possess a regime in which one recovers an effective theory that can be understood as a quantum field theory defined on a fixed (emergent) background space-time. Such a result would be a key consistency check for the theory.

However, the dynamics of matter coupled to spin foams is challenging to explore for different reasons. One of them is the complexity of spin foam quantum gravity itself, e.g. the computation of its fundamental amplitudes. Fortunately, in recent years significant progress has been made to overcome this challenge through numerical means, e.g. by developing an algorithm to explicitly compute the vertex amplitude \cite{Dona:2017dvf,Dona:2019dkf,Gozzini:2021kbt,Dona:2022dxs}, the invention of effective spin foams, which utilize the semi-classical approximation of spin foam amplitudes \cite{Asante:2020qpa,Asante:2020iwm,Asante:2021zzh,Asante:2021phx}, the usage of Lefshetz thimbles making Monte Carlo algorithms applicable to oscillatory spin foam path integrals \cite{Han:2020npv} and the contributions from complex critical points in spin foams with multiple simplices \cite{Han:2021kll}. Despite these advances, performing pure spin foam calculations beyond small triangulations is challenging, not to mention adding matter dynamics on top. Instead, we opt for a restricted, semi-classical spin foam model, called quantum cuboids defined in \cite{Bahr:2015gxa}, which is accessible for numerical simulations and whose amplitudes are derived from the EPRL-FK model defined on a hypercubic 2-complex \cite{Kaminski:2009fm}. Essentially, one can understand this model as a superposition of flat, hypercuboidal (and thus irregular) lattices weighted by spin foam amplitudes. In the last few years, this model gave the first chance to study important questions in spin foam quantum gravity, e.g. renormalization \cite{time-evo,Dittrich:2014ala,Steinhaus:2020lgb} and the spectral dimension \cite{Carlip:2019onx}: In \cite{Bahr:2016hwc,Bahr:2017klw} the renormalization group flow of cuboid spin foams was computed, which showed indications of a UV attractive fixed point and a second order phase transition. These results were later generalized to frusta \cite{Bahr:2017eyi,Bahr:2018ewi}, generalizations of cuboids to cosmological setting allowing for non-vanishing curvature, which confirmed these findings in a more general context \cite{Bahr:2018gwf}. Moreover, in \cite{Steinhaus:2018aav} the spectral dimension, an effective dimension measure derived from a diffusion process, for quantum cuboids was computed, which can be lower than four due to the superposition and scaling of cuboid geometries of different size. Thus, while quantum cuboids are no realistic model of quantum space-time, they pertain features of spin foam models and provide an ideal test scenario to define and study suitable observables. In this work, we will e.g. define an observable, the 2-point correlation of the scalar field in a relational manner \cite{Rovelli:2001bz,Dittrich:2005kc,Tambornino:2011vg}, where we measure both the scalar field correlations as well as their distance encoded in the spin foam state.

As a matter system we consider a free, massive scalar field defined as a lattice field theory on the cuboid spin foam. To do so, we interpret a spin foam configuration as an irregular lattice on which we define scalar lattice field theory via discrete exterior calculus \cite{desbrun_marsden,Calcagni:2012cv}. Then, by superimposing spin foam states we also superimpose lattice field theories defined on these configurations. At the level of the amplitudes, the spin foam is unaffected by the addition of the scalar field, while the latter is sensitive to the spin foam only via the geometry of the spin foam state it is defined on. We consider this a ``minimal coupling'', but it is certainly not free of ambiguities. Due to the non-oscillatory nature of quantum cuboid amplitudes, which even persists into the deep quantum regime \cite{Allen:2022unb}, we can study the whole system with a Markov Chain Monte Carlo algorithm and compute expectation values of observables. Thus, this model gives us the first chance to study the dynamics of a matter theory coupled to a 4d spin foam model and learn valuable first lessons about such systems.

This article is organized as follows: we open in section \ref{sec:regular_scalar_field} with a brief review of free scalar field theory defined on a regular lattice. In section \ref{sec:scalar_irregular} we generalize this field theory to irregular lattices using discrete exterior calculus. Section \ref{sec:spin_foams} briefly introduces spin foam models and the cuboid restriction. We define the coupled system, scalar lattice field theory on quantum cuboids, in section \ref{sec:coupled_system} and describe the Monte Carlo algorithm to compute its expectation values in section \ref{sec:monte_carlo}. The results are detailed in section \ref{sec:results} and we close with a summary and discussion in section \ref{sec:discussion}.

\section{Scalar field theory on a regular lattice} \label{sec:regular_scalar_field}

As an introduction to lattice field theory, we start by briefly discussing the free, massive scalar field defined on a regular, hypercubic lattice. We start from the continuum (Euclidean, Wick-rotated) action:
\begin{equation}
    S[\phi] := \int d^4 x \left( \frac{1}{2}(\partial_\mu \phi)(\partial^\mu \phi) + \frac{M^2}{2} \phi^2   \right) \quad ,
\end{equation}
and define the partition function as:
\begin{equation}
    Z := \int \mathcal{D}[\phi] \; e^{-S[\phi]} \quad . 
\end{equation}
To make this expression well-defined, we regularize it by introducing a discretisation, a regular lattice with lattice constant $a$. On the lattice, the continuum scalar field $\phi(x)$ is replaced by the field $\phi_x$ located at the coordinates $x$ of the vertices of the regular lattice. We discretize the derivatives by approximating them by difference quotients of fields sharing the same edge, but this definition is not unique and different choices differ by lattice artifacts of $\mathcal{O}(a)$. Instead, we consider the lattice Laplacian (by performing an integration by parts and dropping boundary terms due to periodic boundary conditions):
\begin{equation}
    \partial^2 \phi \rightarrow \sum_{\pm \mu} \frac{\phi(x+e_\mu) + \phi(x-e_\mu a) - 2 \phi(x)}{a^2} \quad ,
\end{equation}
where $e_\mu$ denotes a unit vector in $\mu$ direction, and $\pm \mu$ denotes that we are summing over positive and negative directions for each direction $\mu$.

The integral over all points in the continuum is replaced by a sum over all vertices, $\int d^4 x \, \phi(x) \rightarrow a^4 \sum_x \phi_x$, where $a^4$ is the 4-volume of a lattice hypercube. Here $x$ denotes the coordinates of the vertices of the lattice and $\phi_x := \phi(x)$. The functional integral in this discrete setting is defined as a product of regular integrals,  $\int \mathcal{D}[\phi] \rightarrow \prod_x \int d \phi_x$.
Combining all these ingredients, we find the discrete lattice action.
\begin{equation}
    S^{(d)}[\phi] = -\frac{a^2}{2} \sum_{x,\mu} \phi_x \phi_{x+\mu} + \frac{a^2}{2} (8 + a^2 M^2) \sum_x \phi^2_x \quad ,
\end{equation}
where we have used the periodicity of the lattice to rewrite the kinetic part of the action.

This action can be brought into a simpler form, by enumerating the vertices of the lattice and writing the field configuration as a single $N$-dimensional vector $\vec{\phi}$, where $N$ is the number of lattice vertices. Then, we express the scalar field action as the contraction of a matrix $K$ by two vectors of field configurations:
\begin{equation}
    S^{(d)}[\phi] = \frac{1}{2} \phi_n K_{nm} \phi_m \quad ,
\end{equation}
\begin{equation} \label{eq:regular_action}
    K_{nm} = -a^2 \sum_{\mu}(\delta_{n+e_\mu,m} + \delta_{n-e_\mu,m} - 2 \delta_{nm}) + a^4 M^2 \delta_{nm} \quad .
\end{equation}
Given this representation, computing observables such as the 2-point correlation function $\langle \phi_n \phi_m \rangle$ is straightforward and given by the inverse of $K$, $K^{-1}_{nm}$. The correlations of scalar fields at different vertices drops off exponentially with the distance $d(n,m)$ between them i.e., $\langle \phi_n \phi_m \rangle \sim e^{-\frac{d(n,m)}{\xi}}$, where $\xi$ denotes the correlation length. For the massive scalar field, this correlation length is inversely proportional to the mass $M$ of the scalar field, $\xi = M^{-1}$, i.e. the larger the mass the faster the correlations die off as we increase the distance. Conversely, we can estimate the mass of the scalar field from the exponential decay of its correlations.

\subsection{Taking the continuum limit}

In lattice field theory, the lattice plays the role of a regulator and particular care is necessary to make sure that the results are independent of this regularization. To this end, one investigates whether a continuum limit can be taken, i.e. the regulator can be removed. Naively, this limit corresponds to taking $a \rightarrow 0$, but it can only be taken on a second order phase transition in the phase diagram of the theory. Only on such a transition scale invariance is guaranteed and the limit $a \rightarrow 0$ is well defined. Note however, that such a phase transition is not possible for finite volume. The specifics of this procedure go beyond the scope of this article, and we refer to the literature of lattice field theory.

For the free, massive scalar field, the continuum limit can be readily taken. Note that the lattice spacing $a$ also determines the dimension of analytic operations, i.e. dimensionful quantities can be rescaled by (powers of) $a$ to turn both fields and coupling constants into dimensionless quantities. In 4d, we define dimensionless fields $\varphi = \frac{\phi}{a}$, and the scalar field action is written in terms of a new matrix $\tilde{K}$:
\begin{equation}
    S^{(d)}[\varphi] = \frac{1}{2} \varphi_n \tilde{K}_{nm} \varphi_m \quad ,
\end{equation}
\begin{equation}
    \tilde{K}_{nm} = - \sum_{\mu}(\delta_{n+e_\mu,m} + \delta_{n-e_\mu,m} - 2 \delta_{nm}) + \tilde{M}^2 \delta_{nm} \quad ,
\end{equation}
where we have defined the lattice mass $\tilde{M}(a) = a M$. Thus, the scalar field action contains no dependence on the lattice scale besides the lattice mass $\tilde{M}$. Taking the continuum limit corresponds to taking the lattice mass $\tilde{M} \rightarrow 0$, while keeping $M = \frac{\tilde{M}}{a}$ constant. This leads indeed to a second order phase transition, since the correlation length $\tilde{M}^{-1}$, i.e. the physical correlation length expressed in lattice units, diverges. While this is a trivial example of taking the continuum limit, it demonstrates that free, massive scalar field theory has only one free parameter, the mass $M$ of the scalar field. Hence, it should also be a suitable test case for coupling matter to a discrete approach to quantum gravity.

\section{Scalar matter on an irregular discretization} \label{sec:scalar_irregular}

In order to couple scalar lattice field theory to spin foams, we generalize its definition to irregular lattices, which includes both their geometry and combinatorics. Therefore, the construction we present is applicable e.g. to triangulations. To do so, we use discrete exterior calculus \cite{desbrun_marsden,Calcagni:2012cv,Thurigen:2015paa}: the idea is to discretize a continuum action expressed in terms of differential forms by smearing $p$-forms (their duals $(d-p)$-forms) over $p$-simplices (dual $(d-p)$-simplices) respectively. 

On a manifold $\mathcal{M}$ a real scalar field $\phi$ with mass $M$ is a $0$-form ($ \phi \in \Omega^0(\mathcal{M})$) with the action 
\begin{equation}
S = \frac{1}{2}\int_\mathcal{M} d\phi \wedge *d\phi + M^2 \phi *\phi
\end{equation}
where $d\phi$ is the exterior derivative of $\phi$ and $*$ denotes the Hodge dual.

Let us now consider a simplicial discretization $K$ of $\mathcal{M}$. In this discrete setting the role of a $p$-manifold is played by $p$-chains, i.e, formal linear combinations of $p$-simplices \ma{($\sigma_p$)} in $K$ generating the vector space of $p$-chains $c \in C_p(K)$. We write an element of $C_p(K)$ as
\begin{equation}
\ket{c}  = \sum_{\sigma_p \in K_p} \braket{\sigma_p|c} \ket{\sigma_p}.
\end{equation}
Similarly linear forms on $K$ are represented by $p$-cochains $\tilde{c} \in C^p(K)$ written as:
\begin{equation}
\bra{\tilde{c}} = \sum_{\sigma_p \in K_p} \braket{\tilde{c}|\sigma_p}\bra{\sigma_p}.
\end{equation}
This identification between the continuum linear forms and the discrete $p$-cochains assumes that the discretization process involves an integration of $p$-forms over $p$-simplices and in the discrete setting integration is replaced by the evaluation of a cochain on a chain ($\braket{\tilde{c}|c}$). Thus, a $p$-form smeared on a single $p$-simplex may be written as
\begin{equation}
\label{smeared}
\omega(\sigma_p) =  \braket{\omega|\sigma_p} = V_{\sigma_p} \omega_{\sigma_p}=\int_{\sigma_p} \omega \quad .
\end{equation}
Here $\braket{\omega|\sigma_p}$ is the integrated value of the $p$-form $\omega$ on $\sigma_p$ and contains information about the volume of $\sigma_p$, whereas the coefficients $\omega_{\sigma_p}$ are the discrete components of the $p$-form.

Therefore the scalar field can be discretized by smearing on $0$-simplices ($\sigma_0$) as:
\begin{equation}
\phi(\sigma_0) = \braket{\phi|\sigma_0} = \phi_{\sigma_0} \quad ,
\end{equation}
where the vertex volumes are assumed to be trivial in the last equality and the coefficients $\phi_{\sigma_0}$ are the discrete components of the field.

In order to write down the action for this discrete scalar field we need to define the concept of a discrete Hodge dual. Let us then consider the complex $*K$ that is dual to $K$. The complex $*K$ consists of $(d-p)$ cells $\star \sigma_p$ dual to the primal cells $\sigma_p$ with orientation and cellular structure  induced from the orientation and adjacency matrix of $K$.  The duality between the primal and dual simplicial complexes induces a duality between the chains of the primal  ($c \in C_p(K)$) and dual complexes ($ \star c \in C_p(\star K)$) (as well as the co-chains of primal and dual complexes). That is,
\begin{equation}
\ket{c} = \sum_{\sigma_p \in K} c_{\sigma_p} \ket{\sigma_p} \longleftrightarrow \bra{\star c} = \sum_{\sigma_p \in K} c^*_{\sigma_p} \bra{\star\sigma_p}
\end{equation}
Along with the identification between $p$-forms and cochains on $K$, this allows the definition of the discrete Hodge dual of a $p$-form as
\begin{equation}
*\omega  = \star \omega \in *\Omega^p(\mathcal{M}) \sim \Omega^{d-p}(\star K) \sim C^{d-p}(\star K)
\end{equation}
 At the level of coefficients the duality reads: 
\begin{equation}
\label{hodge-star} 
\phi_{\sigma_p} \overset{\star}{\mapsto} (\star \phi)_{\star \sigma_{p}} = \phi_{\sigma_p}^{*}.
\end{equation}
For the smeared $p$-forms this implies:
\begin{equation}
\star \left(\frac{\braket{\omega|\sigma_p}}{V_{\sigma_{p}}} \right ) = \frac{\braket{\star \sigma_p|\omega}}{V_{\star \sigma_p}}.
\end{equation}
With these definitions, we can define a pairing of the $p$-chains on $K$ and its dual on $\star K$ as $\braket{\sigma_p | \star \sigma_{p'}} = \delta_{p,p'}$ and a completeness relation $\sum_{p \in K} \ket{\sigma_p} \bra{\star \sigma_p} = \text{id}$ \cite{Thurigen:2015paa}.

In order to define an action for the discrete field we need the definition of the discrete exterior derivative. The discrete exterior derivative is defined in order to satisfy a discrete version of the Stokes' theorem $\braket{\mathrm{d}\omega|c} = \braket{\omega|\delta c}$, such that the exterior derivative $\mathrm{d}$ is the natural dual of the boundary operator $\delta$ as in the continuum. The boundary operator is defined by its action on a p-simplex with vertices $[v_1...v_p]$:
\begin{equation}
    \delta \sigma_p = \sum_i^p (-1)^i [v_1...(v_i)...v_p] = \sum_{\sigma_{p-1} \in \sigma_{p}} \text{sgn}(\sigma_{p-1},\sigma_{p}) \sigma_{p-1}.
\end{equation}
where the first equality is the alternating sum over all $(p-1)$-simplices obtained by dropping a vertex in $\sigma_p$ (the parenthesis denotes the vertex that has been dropped) and the second equality is just a compact rewrite.

Thus we can write the discrete analog of the exterior derivative of a $p$-form $\omega$ as:
\begin{equation}
\label{exterior-derivative}
 \mathrm{d}\omega(\sigma_{p+1}) = \int_{\sigma_{p+1}} \mathrm{d}\omega = \int_{\delta \sigma_{p+1}} \omega = \omega(\delta \sigma_{p+1}) = \sum_{\sigma_{p} \in \sigma_{p+1}} \text{sgn}(\sigma_p,\sigma_{p+1}) \omega(\sigma_p)   
\end{equation}

We now have all the ingredients to write the discrete action for the scalar field. It takes the form:
\begin{equation}
S_D  = \frac{1}{2} \left( \braket{\mathrm{d}\phi|\mathrm{d}\phi} + M^2 \braket{\phi | \phi} \right) = \frac{1}{2} \left( \braket{\star d\star d\phi|\phi} + M^2 \braket{\phi | \phi} \right),
\end{equation}
where the second equality arises after integration by parts and setting the boundary term to zero, e.g. when using periodic boundary conditions. Interested readers will find further details in \cite{desbrun_marsden,Thurigen:2015paa}.

Using equations (\ref{smeared}),(\ref{hodge-star}) and (\ref{exterior-derivative}) repeatedly\footnote{For example, the potential term follows from $\braket{\phi|\phi} = \sum_{\sigma_0} \braket{\phi|\sigma_0} \braket{\star \sigma_0| \phi} = \sum_{\sigma_0} \phi_{\sigma_0} (V_{\star \sigma_0} \phi_{\sigma_0})$.} the discrete action can be written as
\begin{equation}
S_D =  \sum_{\sigma_0} \phi(\sigma_0) \sum_{\sigma_1 \supset \sigma_0} \frac{V_{\star \sigma_1}}{V_{\sigma_1}}  \text{sgn}(\star\sigma_1, \star \sigma_0) \sum_{\sigma_{0}' \subset \sigma_1} \text{sgn}(\sigma'_0,\sigma_1) \phi(\sigma'_0) + \frac{M^2}{2} \sum_{\sigma_0} V_{\star \sigma_0} \phi(\sigma_0)^2 \quad .
\end{equation}
In 4d, the case we are interested in, $V_{\sigma_1}$ is the length of the edge $\sigma_1$, whereas $V_{\star \sigma_1}$ is the volume associated to its dual $3$-cell. $V_{\star \sigma_0}$ denotes the $4$-volume of the $4$-cell dual to a vertex. After a little rearranging we can write this is in the form $S_D = \frac{1}{2} \sum_{\sigma^i_0, \sigma^j_0} \phi(\sigma_0^i) \, K_{ij} \, \phi(\sigma^j_0)$ where
\begin{equation} \label{eq:irregular_action}
K_{ij } =  \begin{cases} 
       \sum_{\sigma_1 \supset \sigma_0^i}  \frac{V_{\star \sigma_1}}{V_{\sigma_1}}  \text{sgn}(\star\sigma_1, \star \sigma_0) \text{sgn}(\sigma_0,\sigma_1) + M^2 \, V_{\star \sigma_0} &  i == j \\
       \\
        \frac{V_{\star \sigma_1}}{V_{\sigma_1}}  \text{sgn}(\star\sigma_1, \star \sigma_0^i) \text{sgn}(\sigma_0^j, \sigma_1) & i \neq j \,\, \text{and} \,\, \sigma_0^i, \sigma_0^j \subset \sigma_1 \quad .
   \end{cases}
\end{equation}
Note that the different components of $K_{ij}$ associated to the Laplacian and the potential have the same scaling behaviour in terms of lengths as the scalar field theory action discretized on a regular lattice in section \ref{sec:regular_scalar_field}\footnote{Careful readers might notice differences to the Laplacian defined in previous work by one of the authors \cite{Steinhaus:2018aav}. There it is defined on the dual complex, and only the action of the Laplacian on a test field is considered, not the action integrated over the entire complex.}. Indeed, \eqref{eq:irregular_action} simplifies to \eqref{eq:regular_action} if defined on a regular lattice, where the dual lattice is simply given by the same (shifted) lattice.

The definition of the scalar field action \eqref{eq:irregular_action} can be used for general cellular complexes, but its definition is not unique, e.g. it is a priori not clear whether to smear the scalar field over vertices of the complex or over vertices of the dual complex. Additionally, there is freedom in constructing the dual lattice and thus defining the dual volumes.

In the next section we briefly introduce spin foam models and the restricted model we will couple with scalar lattice field theory.

\section{Restricted spin foam models} \label{sec:spin_foams}

Coupling a lattice field theory to spin foam models and examining this system numerically in full generality is a daunting task. Thus, in order to get a first insight into such spin foam-matter systems and to determine which methods are suitable to investigate such models, we will study a spin foam path integral restriced to so-called ``quantum cuboids'' \cite{Bahr:2015gxa}. In this section, we give a brief introduction to spin foam quantum gravity \cite{alexreview} and define this restricted model.

Similar to lattice field theories, spin foam models are defined on a discretisation, more precisely a 2-complex, which is a collection of vertices $v$, edges $e$ and faces $f$. Frequently this 2-complex is chosen dual to a triangulation, but it can be more general \cite{Kaminski:2009fm}; here we choose it to be dual to a (hyper)cubulation. The discrete geometry of the spin foam is then encoded in group theoretic data assigned to the 2-complex. To each face we assign an irreducible representation of the underlying symmetry group, while we assign an invariant tensor, called an intertwiner, to each edge.

Let us explain these concepts for the concrete case of $\text{SU}(2)$ BF theory \cite{bf-theory}: an edge $e$ in the 2-complex is shared by several faces $f \supset e$. Each face $f$ carries an irreducible representation $j_f$, where we assign to the edge $e$ a vector space $V_{j_f} \otimes V^*_{j_f}$ for each $f \supset e$. Thus, each edge $e$ is endowed with a tensor product of representation spaces from the faces by which it is shared, where we assign the various representation spaces and their duals to the source and target vertex of the edge\footnote{The details depend on the fiducial orientations of edges and faces, but do not affect the results and are not relevant for the calculations / simulations in this article.}. However, not the entire vector space is permitted: to each edge we assign a projector onto the subspace invariant under the action of $\text{SU}(2)$, called the Haar projector:
\begin{equation}
    \mathcal{P}^{j_1,\dots,j_n}_{m_1,\dots,m_n,n_1,\dots,n_n} := \int_{\text{SU}(2)} dg \, D^{j_1}_{m_1 n_1} (g) \dots D^{j_n}_{m_n n_n}(g) \quad ,
\end{equation}
where $D^j$ denote Wigner matrices in representation $j$ and magnetic indices $m,n$ with $-j \leq m,n \leq j$. These projectors can be expressed as a (finite) sum over orthonormal intertwiners $\iota_d$
\begin{equation}
    \mathcal{P}^{j_1,\dots,j_n}_{m_1,\dots,m_n,n_1,\dots,n_n} = \sum_d |\iota_d \rangle \langle \iota_d | \quad .
\end{equation}
As an example, if the edge is 4-valent as in a 4d triangulation, the intertwiner basis elements are labelled by an additional $\text{SU}(2)$ spin arising from expanding the 4-valent intertwiner into two 3-valent ones using recoupling theory. Given this projector property and the fact that an edge connects two vertices, we associate one intertwiner to each vertex of the edge. Then, at the vertex the intertwiners get contracted according to the combinatorics of the vertex. This spin network evaluation defines the vertex amplitude.

Additionally, spins and intertwiners permit a quantum geometric interpretation: we think of an edge to be dual to a polyhedron, with as many faces as the valency of the edge. The areas of the faces are given by the spins associated to them, while the intertwiner encodes the shape. For a tetrahedron, an intertwiner in the orthonormal basis is given by five parameters, four areas and one intertwiner label. The last label corresponds to the area of a parallelogram inside the tetrahedron according to the recoupling scheme. This does not uniquely determine the shape of the tetrahedron, which would require six parameters, e.g. six edge lengths. Thus, this is often referred to as a quantum tetrahedron \cite{Baez:1999tk}, and similar arguments also apply to more general polyhedra.

Modern 4d spin foam models like the Euclidean EPRL-FK model \cite{eprl,FK} we are considering in this article use BF theory as their starting point. In order to break the topological nature of BF theory and arrive at a gravitational theory, simplicity constraints are imposed. To each polyhedron we assign a time-like normal\footnote{We enforce simplicity constraints by demanding that all bivectors in a polyhedron are orthogonal to a common 4d vector. Then each bivector is simple, i.e. a wedge product of vectors. Note that we refer here to the intrinsic Minkowski metric (not a global one). We refer to \cite{Conrady:2010kc} for a more detailed discussion and explanation of this construction.}, such that - as seen from 4d Minkowski space - the polyhedron is entirely space-like. This normal singles out a subgroup of the full symmetry group ($\text{Spin}(4)$ in Euclidean and $\text{SL}(2,\mathbb{C})$ in the Lorentzian signature), which stabilizes the normal. This subgroup is isomorphic to $\text{SU}(2)$, and we construct states of the 4d spin foam model by embedding $\text{SU}(2)$ states into states of the 4d symmetry group. This leads to restrictions on the representation labels depending on the Barbero-Immirzi parameter $\gamma$ compared to 4d BF theory. Furthermore, on the boundary we can connect to the kinematical Hilbert space of loop quantum gravity via projected spin network states \cite{Livine:2002ak,Dupuis:2010jn}.

For the Euclidean EPRL model for $\gamma < 1$, the symmetry group $\text{Spin}(4) \cong \text{SU}(2) \times \text{SU}(2)$ and its representations $(j^+,j^-)$ are labelled by two $\text{SU}(2)$ representations $j^\pm \in \frac{\mathbb{N}}{2}$. They are related to the same $\text{SU}(2)$ representation as follows:
\begin{equation}
    j^\pm = \frac{j}{2} (1 \pm \gamma) \quad ,
\end{equation}
where $j \in \frac{\mathbb{N}}{2}$. This immediately requires $\gamma \in \mathbb{Q}$ for this map to be non-trivial, and is considered a pathology of the Riemannian model, which is absent in the Lorentzian theory.

To implement this construction for the intertwiners, we define a boost map $Y_\gamma$ consisting of two parts. For an $\text{SU}(2)$ intertwiner $\iota$, we isometrically embed each vector space $V_j$ into the unique component appearing in the Clebsch-Gordan decomposition of $V_{j^+,j^-} \simeq V_{j^+} \otimes V_{j^-}$. We call this map $\beta^\gamma_j$ as it explicitly depends on $\gamma$. Note that the resulting tensor is not necessarily a $\text{SU}(2) \times \text{SU}(2)$ intertwiner; we thus act with the Haar projector $\mathcal{P}$. For an $N$-valent intertwiner $Y_\gamma$ reads:
\begin{align}
    Y_\gamma : \text{Inv}(V_{j_1} \otimes \dots \otimes V_{j_N}) & \rightarrow \text{Inv}(V_{j_1^+,j_1^-} \otimes \dots \otimes V_{j_N^+,j_N^-}) \nonumber \\
    Y_\gamma & := \mathcal{P} \circ (\beta^\gamma_{j_1} \otimes \dots \otimes \beta^\gamma_{j_N}) \quad .
\end{align}
Eventually, the vertex amplitude $\mathcal{A}_v$ is defined as the contraction of intertwiners associated with the vertex $v$, also called the vertex trace.
\begin{equation}
    \mathcal{A}_v := \text{Tr}_{e \supset v} (Y_\gamma(\iota_e)) \quad .
\end{equation}

To complete the definition of the model, we must assign amplitudes to the edges $e$ and faces $f$. The edge amplitude $\mathcal{A}_e$ is given by the norm of the associated intertwiner $\iota_e$:
\begin{equation}
    \mathcal{A}_e(\iota_e) = \frac{1}{|| Y_\gamma(\iota_e)||^2} \quad .
\end{equation}
To the faces we assign the dimension of the representations $j^+,j^-$ up to an exponent $\alpha$
:
\begin{equation}
    \mathcal{A}_f(j_f;\alpha) := ((2 j^+_f + 1)(2 j^-_f +1))^\alpha \quad .
\end{equation}
We introduce this parameter $\alpha$ by hand to reflect an ambiguity in defining the face amplitude of the model. Common choices are either the SU$(2)$ representation $2 j_f+1$ or the SU$(2) \times $SU$(2)$ representation $(2j^+_f +1)(2j^-_f +1)$. The former choice has the advantage of being invariant under face and edge subdivisions \cite{Bianchi:2010fj}. Therefore, we understand the parameter $\alpha$ as a modification of the path integral measure. In the quantum cuboid model we introduce below, these face amplitudes can be rewritten as the volume of 4d  hypercuboids to a power of a multiple of $\alpha$. This is similar to path integral measure choices proposed in quantum Regge calculus \cite{Hamber:1997ut}.
The choice of $\alpha$ has a profound impact on the scaling behavior of the amplitudes and played a crucial role in the coarse graining \cite{Bahr:2016hwc,Bahr:2017klw} and the spectral dimension \cite{Steinhaus:2018aav} of restricted 4D spin foams models. We will also see that it has a significant impact on the coupled matter spin foam system.

Finally, the spin foam partition function is defined as:
\begin{equation}
    Z = \sum_{j_f,\iota_e} \prod_f \mathcal{A}_f \prod_e \mathcal{A}_e \prod_v \mathcal{A}_v \quad .
\end{equation}
where we sum over all spin foam states given by assignments of representations and intertwiners.

\subsection{Cuboid spin foams}

Spin foam amplitudes are well-defined, yet the complexity of the theory makes actual calculations demanding. This is due to the sheer number of configurations one has to consider even for small 2-complexes as well as the challenge to compute its fundamental amplitude beyond asymptotic methods. To at least explore a subset of the full spin foam path integral, restricted semi-classical models were introduced in \cite{Bahr:2015gxa,Bahr:2017eyi,Assanioussi:2020fml}, and simplify the theory in different ways:
\begin{itemize}
    \item \textit{Hypercubic combinatorics}: We choose the 2-complex to be dual to a 4d cubulation. This has the advantage that the combinatorics is regular, e.g. each face is bounded by four edges and each edge is 6-valent etc. The initial motivation for this choice in \cite{Bahr:2015gxa} was to simplify renormalization / coarse graining \cite{Bahr:2016hwc,Bahr:2017klw,Bahr:2018gwf}, but here it will be helpful to relate our model to lattice field theory.
    \item \textit{Intertwiner restriction}: Spin foam models with hypercubic combinatorics are generically more complicated than their counterparts defined on (dual) triangulations. This is due to the fact that more data are required to specify the building blocks, e.g. a hypercubic vertex has eight edges and 24 faces compared to the five edges and ten faces of a vertex dual to a 4-simplex. This is counteracted by only allowing specific coherent intertwiners that are peaked on a cuboid \cite{Bahr:2015gxa} or a frustum shape \cite{Bahr:2017eyi}. These choices also restrict the representations, reducing the complexity of the model further. 
    \item \textit{Semi-classical amplitude}: The first numerical computation of the vertex amplitude for hypercubic combinatorics has only recently been achieved beyond smallest spins \cite{Allen:2022unb}, however at great numerical costs. Thus it is currently out of reach to explore larger 2-complexes using the full amplitude; instead we will use its semi-classical approximation derived in \cite{Bahr:2015gxa,Bahr:2017eyi}.
    \item \textit{Continuous variables}: The semi-classical approximation is typically derived in the large $j$-limit for coherent boundary data, where the representation labels of the vertex amplitude are uniformly scaled up. In this limit, the discreteness of representations is barely noticeable and we approximate them as continuous variables.
\end{itemize}
Most of these assumptions are drastic simplifications, restricting the full spin foam model to a subset of the gravitational path integral that is more readily accessible by numerical means. Note that this does not fix a background geometry a priori, as we sum over all configurations permitted by these restrictions. However, as we will see in this article, the system might be dominated by particular configuration (plus fluctuations around it), which one can interpret as the emergence of a dynamical background.
Clearly, the restricted models do not capture all properties of spin foams, and these assumptions must be eventually removed in future research.

\begin{figure}
    \centering
    \includegraphics[width = 0.3\textwidth]{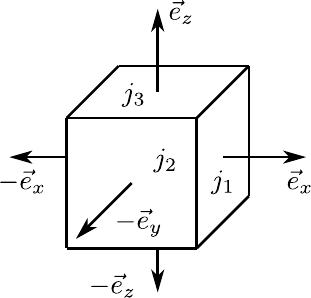}
    \caption{The shape of a cuboid is encoded in the areas of its faces $j_i$ and the the outward pointing normal vectors.}
    \label{fig:cuboid}
\end{figure}

The restriction of intertwiners is the final piece left to define the concrete restricted model. We choose the intertwiners to be coherent Livine-Speziale intertwiners \cite{Livine:2007ya} peaked on the shape of cuboids, dubbed cuboid intertwiners. Coherent $\text{SU}(2)$ intertwiners (and also the boosted intertwiners) are parametrised by spins $j_i$ and normal vectors $\vec{n}_i \in S^2$ assigned to their faces, which agree with the areas and outward pointing normals of the classical polyhedron on which they are peaked. For cuboids, opposite faces carry the same area and opposite normals are anti-parallel, while each normal is perpendicular to the normals of its four adjacent faces, see also fig. \ref{fig:cuboid}. We define the intertwiner as follows:
\begin{equation}
    \iota_{j_1,j_2,j_3} := \int_{\text{SU}(2)} dg \, g \triangleright \bigotimes_{i = 1}^3 |j_i,\vec{e}_i \rangle \otimes |j_i,-\vec{e}_i \rangle \quad .
\end{equation}
The states $|j_i,\vec{e}_i\rangle \in V_j$ denote $\text{SU}(2)$ Perelomov coherent states \cite{perelomov}, i.e. maximum (or minimum) weight states on which we act with $\text{SU}(2)$ representation matrices. The state $|j,\vec{n}\rangle$ is defined as $g_{\vec{n}} \, \triangleright \, |j,j\rangle =: g_{\vec{n}} \, \triangleright \, |j,\vec{e}_z\rangle$, where $g_{\vec{n}}$ is the group element encoding the rotation $\vec{e}_z \rightarrow \vec{n}$. By $\vec{e}_i$ we denote the unit vectors for Cartesian coordinates. The group integration, also called group averaging, ensures the invariance of the intertwiner under $\text{SU}(2)$ transformations.

Boosting these intertwiners for $\gamma < 1$ is straightforward: each coherent state $|j,\vec{n}\rangle$ gets mapped to a tensor product of coherent states $|j^+,\vec{n}\rangle \otimes |j^-,\vec{n}\rangle$ for the representations $j^\pm$. The $\text{SU}(2) \times \text{SU}(2)$ group averaging is then implemented as two group integrations, one for all ``$+$'' and one for all ``$-$'' states. Thus, the resulting intertwiner is a tensor product of two $\text{SU}(2)$ intertwiners, one for ``$+$'' and one for ``$-$'' labels, peaked on the same coherent data $\{\vec{n}\}$. See \cite{FrankEPRL,Conrady-Freidel} for a more detailed derivation and explanation.

\subsubsection{Semi-classical approximation of amplitudes}

\begin{figure}
    \centering
    \includegraphics[width = 0.4 \textwidth]{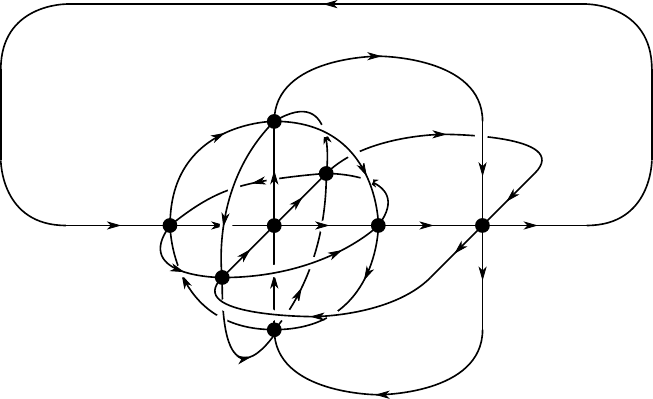} \; \;
    \includegraphics[width = 0.4 \textwidth]{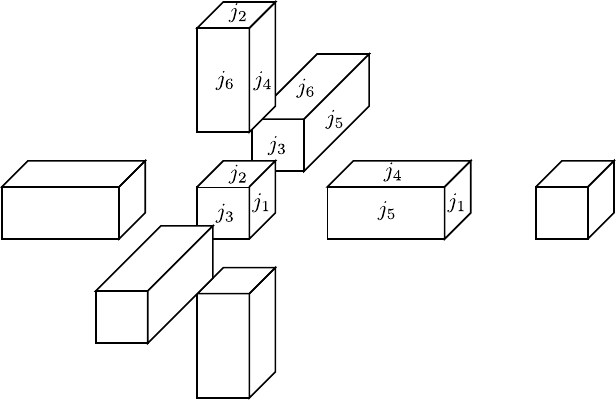}
    \caption{\textit{Left:} Spin network graph for a spin foam vertex with the combinatorics of a dual 4d hypercube. \textit{Right:} The 3d boundary corresponding to a hypercuboid and the area data matching the spin foam vertex amplitude.}
    \label{fig:vertex_snw}
\end{figure}

The vertex amplitude for cuboid spin foams is given by the evaluation of the spin network in fig. \ref{fig:vertex_snw}. For coherent boundary data, it reads:
\begin{align}
    \mathcal{A}_v := & \int_{\text{SU}(2)^8} \prod_{a=1}^8 dg_a^+  \prod_{a<b} (\langle - \vec{n}_{ba} |(g^+_b)^{-1} g^+_a |\vec{n}_{ab} \rangle )^{2 j^+_{ab}} \; \times \nonumber \\
    \times & \int_{\text{SU}(2)^8} \prod_{a=1}^8 dg_a^- \prod_{a<b} (\langle - \vec{n}_{ba} |(g^-_b)^{-1} g^-_a |\vec{n}_{ab} \rangle )^{2 j^-_{ab}} \nonumber \\
    =: & \mathcal{A}_v^+ \mathcal{A}_v^- \quad .
\end{align}
We denote the intertwiners in the amplitude by $a,b$, and have one group integration for each. Then, to each link connecting two intertwiners, we assign an inner product of coherent states labeled by their respective normal vectors: $\vec{n}_{ab}$ denotes the normal vector to the face $ab$ belonging to the polyhedron $a$ and $\vec{n}_{ba}$ vice versa. Moreover, we have used the property of coherent states (as maximum weight states) $\langle j,\vec{n}_1| \cdot |j, \vec{n}_2 \rangle = (\langle \vec{n}_1| \cdot |\vec{n}_2 \rangle)^{2j}$, where the latter expression is given in the fundamental spin $\frac{1}{2}$ representation. For $\gamma<1$ the amplitude factorizes over the ``$+$'' and ``$-$'' labels, such that we can write it as the product of two $\text{SU}(2)$ BF theory vertex amplitudes for specific representations.

These group integrations are highly oscillatory integrals for large $j_{ab}^\pm$. Thus, the vertex amplitude is approximated by performing a stationary phase approximation of the group integrals: to do so, the inner products of $\text{SU}(2)$ coherent states are exponentiated to define an action:
\begin{equation}
    S^\pm := 2 \sum_{a<b} j^\pm_{ab} \, \ln \langle - \vec{n}_{ba} |(g^\pm_b)^{-1} g^\pm_a |\vec{n}_{ab} \rangle \quad .
\end{equation}
The vertex amplitude is dominated by the critical points of $S^\pm$ if all spins $j^\pm_{ab}$ are large. These critical and stationary points are derived by varying the action with respect to the dynamical variables, here group elements, and enforcing that $\Re S^\pm = 0$. Variation with respect to the group elements enforces closure of the coherent data $\sum_{b \neq a} j^\pm_{ab} \, \vec{n}_{ab} = 0 \; \forall a$, while the reality conditions determine how the polyhedra are glued together. Again see \cite{Bahr:2015gxa} for more details.

With the critical and stationary points found, we approximate the vertex amplitude $\mathcal{A}^\pm_v$ as follows:
\begin{equation}
    \mathcal{A}^\pm_v \approx \sum_{g_a^{(c)}} \frac{(2 \pi)^{\frac{21}{2}}}{\sqrt{\det (-H)}|_{g_a^{(c)}}} \exp(S^\pm |_{g_a^{(c)}}) \quad ,
\end{equation}
where we sum over the critical point solutions $g^{(c)}_a$. The action $S^\pm$ and the determinant of the Hessian matrix $H$ are evaluated on these solutions. We get one factor of $\sqrt{2 \pi}$ for each integration; we integrate over seven $\text{SU}(2)$ group elements (one integration can be absorbed due to gauge invariance), which are each three-dimensional. For cuboid boundary data, this formula simplifies drastically \cite{Bahr:2015gxa}:
\begin{equation}
    \mathcal{A}^\pm_v \approx \left (\frac{1 \pm \gamma}{2} \right)^{\frac{21}{2}} \mathcal{B}_v \quad , \quad  \mathcal{B}_v = \left(\frac{2}{16 \pi^2}\right)^7 (2 \pi)^{\frac{21}{2}} \left(\frac{1}{\sqrt{ \det (-H)}} + \text{cc.}\right) \quad .
\end{equation}
The action $S^\pm$ has two critical points with a multiplicity of $2^7$ due to a discrete symmetry. We also obtain one factor of $(16 \pi^2)^{-1}$ for each $\text{SU}(2)$ integration. Note that the action evaluated on the critical points, which corresponds to the Regge action \cite{regge} associated with the 4d polytope, vanishes. This has two important consequences: firstly, for larger complexes all deficit angles identically vanish such that we only find flat geometries. Essentially we are gluing together flat geometries in a flat way, corresponding to a cubulation of flat space-time. Secondly, due to the vanishing of the Regge action the vertex amplitude is not oscillatory and determined by the scaling behavior encoded in the determinant of the Hessian. Thus, we can also pull the dependence on the Immirzi parameter $\gamma$ in front of the amplitude. The non-oscillating behavior of the cuboid amplitude even persists in the quantum regime \cite{Allen:2022unb}.

Due to the symmetry of the cuboid intertwiners, i.e. the fact that opposite sides of the 3d cuboids carry the same spin, the amplitude $\mathcal{B}_v$ is a function of six $\text{SU}(2)$ spins. This is peculiar since a 4d hypercuboid is determined by just four edge lengths; this mismatch comes about as follows. Each 3d cuboid is specified by three spins, which determine the six areas of its faces. As long as these areas are non-vanishing, we can uniquely translate three areas into three edge lengths for each cuboid. Depending on the choice of six spins, these edge lengths may however not agree, such that the faces shared between two cuboids have the same area but their shapes do not match. In larger 2-complexes, this can lead to torsion effects where a loop may not close due to the mismatch of lengths in neighbouring hypercuboids \cite{Bahr:2015gxa}.

In the following we consider the subset of spin configurations for which the edge lengths, as seen by different cuboids, agree and shape matching is ensured. The non-shape matching is related to the so-called volume simplicity constraint, which is one of several simplicity constraints imposed at the discrete level. In a classical 4-simplex, expressed as a bivector geometry, the volume of this 4-simplex is spanned by any pair of bivectors associated to triangles that only share one vertex. Again, for a 4-simplex this constraint is automatically satisfied once all other simplicity constraints are implemented, which is the reason why it is not enforced in the original EPRL model. However, this no longer holds for more general 2-complexes  \cite{Belov:2017who,Bahr:2017ajs,Assanioussi:2020fml}, generically giving rise to non-matching shapes \cite{Langvik:2016hxn,Dona:2020yao}.

In \cite{Bahr:2015gxa} also the semi-classical expressions for the edge and face amplitudes were derived. The edge amplitude for the coherent cuboid intertwiners is a simple functions of its three spins:
\begin{equation}
    \mathcal{A}_e(j_1,j_2,j_3) \sim \frac{(j_1+j_2)(j_2+j_3)(j_1+j_3)}{8 (1-\gamma^2)^{-\frac{3}{2}}} \quad .
\end{equation}
Similarly, the face amplitude reads:
\begin{equation}
    \mathcal{A}_f \sim j_f^{2 \alpha} \quad .
\end{equation}

As a final point, due to the regular combinatorics of the hypercubic spin foams, we can combine vertex, edge and face amplitudes into a single amplitude associated to the vertex. Each edge is shared by two vertices, while each face contains four vertices; hence we define:
\begin{equation}
    \hat{\mathcal{A}}_v := \prod_{f \supset v} \mathcal{A}^{\frac{1}{4}}_f \prod_{e \supset v} \mathcal{A}^{\frac{1}{2}}_e \mathcal{A}_v \quad .
\end{equation}

\subsubsection{From spins to lengths}

Let us briefly discuss the transition from spins to lengths. Firstly, we must restrict the spin configurations of the hypercuboid such that the areas of its rectangles arise from four edge lengths. To do so, we express two of the spins per hypercuboid as functions of the remaining four spins. However, since we are originally integrating over all spins in the path integral, we effectively gauge fix the redundant spins and account for this by including a Fadeev-Popov determinant. In the second step, we obtain a Jacobian for the change of variables from the four remaining spins to four edge lengths. Both steps are explained in detail in \cite{Bahr:2017klw}.

Eventually, due to the regular combinatorics, we combine face, edge and vertex amplitudes as well as Jacobian and Fadeev-Popov determinant into a common amplitude that we associate to the vertices $\hat{\mathcal{A}}_v$, which is a function of four lengths of the hypercuboid and the parameter $\alpha$\footnote{The Immirzi parameter only appears as a factor in the partition function and thus does not affect observables. This is due to the vanishing of the Regge action.}. This vertex amplitude is a homogeneous function of degree $24\alpha - 14$:
\begin{equation}
    \hat{\mathcal{A}}_v (\{\lambda l_i\}) = \lambda^{24 \alpha - 14} \hat{\mathcal{A}}_v (\{l_i\}) \quad .
\end{equation}
The spin foam partition function is then given by:
\begin{equation}
    Z_{\text{SF}} = \int \prod_i dl_i \prod_v \hat{\mathcal{A}}_v (\{l_i\}) \quad .
\end{equation}
In this partition function we integrate over all possible lengths assignments to the 2-complex, which are weighted by the spin foam amplitudes.
Note that the amplitude $\hat{\mathcal{A}}_v$ is positive and non-oscillatory. Thus, after introducing upper and lower cut-offs for the lengths $l_i$, we can use the partition function to define a probability distribution suitable for Markov Chain Monte Carlo techniques. This probability distribution then determines the likely length configurations of the spin foam.

\section{Scalar matter coupled to spin foams} \label{sec:coupled_system}

In the previous sections \ref{sec:scalar_irregular} and \ref{sec:spin_foams} we have defined the two models individually that we couple in this work, i.e. a free, massive scalar field defined on a general, irregular lattice and cuboid spin foams, which describe superpositions of discrete flat space-times. Our idea for coupling both theories is as follows: the discrete scalar field action is a function of various (dual) volumes of the underlying irregular lattice. Given a specific spin foam state, here a configuration of lengths $\{l_i\}$, we define the volumes in the scalar field action as functions of these lengths, such that the spin foam describes the space-time the scalar field is defined on\footnote{This ansatz contains the implicit assumption that the matter action / theory is diagonalized by this spin foam representation.}. A similar idea is also used in \cite{oriti-pfeiffer} to couple Yang-Mills theory to the Barrett Crane spin foam model \cite{Barrett-Crane}. Then, we integrate / sum over all field configurations and lengths configurations to define the partition function.

In this setup, we solely modify the matter part of the system without altering the gravitational side
in analogy to general relativity. There, the Einstein Hilbert action remains unchanged, while the matter action has an explicit dependence on the metric in contrast to its definition in flat space-time. Thus we will refer to our ansatz as ``minimal coupling'': the spin foam amplitudes remain unchanged, while we include an explicit dependence on the geometry, here given as a function of the edge lengths, in the matter action.
Explicit modifications of the gravitational part are conceivable, but they are not subject of this work. However, the coupling of matter and spin foams is not uniquely defined. One important choice concerns ``where'' the matter degrees of freedom are placed: the actual discretisation (here the cubulation) or the dual 2-complex. We can either decide to place the scalar field on the vertices of the cubulatation or on the vertices of the 2-complex, dual to each 4d hypercuboid. In the example we are considering here, where the combinatorics of the dual 2-complex is the same as the lattice and we are imposing periodic boundary conditions, we do not expect this choice to have a major impact on the dynamics. Nevertheless, these choices lead to different theories because in general the matter theories are not dual to themselves. Moreover, if the space-time possesses a boundary, this choice affects the boundary Hilbert spaces we associate to them.

In this work, we choose to place the scalar field on the vertices $\sigma_0$ of the cubulation, and infer the volumes from the lengths describing the spin foam. Volumes of the lattice itself, e.g. the length $V_{\sigma_1}$, are computed directly from the lengths. Analogously, we directly compute the dual volumes from the dual lengths; the lattice dual to a hypercuboidal lattice is again a hypercuboidal lattice. Since the spin foam model we are considering is restricted to hypercuboidal geometries, we simply place the vertices of the dual lattice in the center of the hypercuboids. Thus, the dual lattice is again a hypercubulation with the dual edges connecting the centers of neighbouring hypercuboids, from which we straightforwardly compute the dual volumes.
All of these definitions are simple and straightforward due to the rigid right angles of the geometries given by cuboids.
Note that in general the construction of a dual lattice is not unique, which thus influences the definition of the matter action, see e.g. \cite{Hamber:1985gx} for further details.
Finally, we define the partition function of the coupled system as:

\begin{equation}
\label{partition-function}
    Z = \int \prod_i dl_i \, \int \prod_{a \in \sigma_0} d \phi_a \prod_{v \in \star \sigma_0} \hat{\mathcal{A}}_v(\{l_i\}_{i \in \sigma_4}) \; \exp(\sum_{a,b} \phi_a \, K_{ab}(M,\{l_i\}) \, \phi_b) \quad .
\end{equation}
Here we denote the edges of the cubulation by $i$ and the vertices of the cubulation by $a,b$. $\sigma_4$ are the hypercuboids and $\sigma_0$ are the vertices of the cubulation. $M$ denotes the mass of scalar field.

\subsection{Qualitative behavior from regular lattices} \label{sec:qualitative_results}

It is difficult to extract qualitative behaviors from the composite system by analytical methods alone. To gain some initial insight into the qualitative features of the model, we set all lengths of the spin foam equal (yet dynamical), such that we consider a superposition of regular lattices of different sizes. As discussed above, the spin foam vertex amplitude has a scaling behavior of $l^{24 \alpha -14}$. Additionally, for fixed $l$ we perform the Gaussian integrations of the scalar field, which leads to a factor in the integrand of the following form (in four dimensions and $N$ lattice sites):
\begin{equation}
    \frac{1}{\sqrt{l^{2N} \left(\sum_{i = 1}^{N} a_i l^{2i} M^{2i} \right)}} \quad .
\end{equation}
$a_i$ denote numerical factors which are not relevant for our argument here. Thus, the probability distribution for the length $l$ is given by:
\begin{equation}
    \frac{1}{Z} \frac{l^{N (24 \alpha -14)}}{\sqrt{l^{2N} \left(\sum_{i = 1}^{N} a_i l^{2i} M^{2i} \right)}} \quad .
\end{equation}
Thus, we observe two counteracting behaviors of the scalar field and the spin foam (for sufficiently large $\alpha$): the scalar field leads to a polynomial suppression of large lengths, which gets stronger as we increase the mass $M$. Conversely, the spin foam amplitude eventually favors larger lengths as we increase $\alpha$. Thus we can readily identify two qualitative regimes of the model:
\begin{itemize}
    \item Finite $M$ and small $\alpha$: $\alpha$ is too small to counteract the suppression of large lengths by the scalar field. Thus, the amplitude diverges as we approach $l \rightarrow 0$ and the smallest lengths dominate. To explore this regime in the actual simulations we introduce a lower cut-off on the lengths. Note that such small lengths are outside the validity of the semi-classical approximation of spin foam models.
    \item Finite $M$ and large $\alpha$: $\alpha$ is so large that it dominates over the suppression of large lengths by the scalar field. Then, the amplitude diverges for $l \rightarrow \infty$ and largest lengths dominate. Thus we introduce also an upper cut-off in the simulations.
\end{itemize}
Note that for fixed lattice size the values of $\alpha$ when either the smallest or the largest lengths dominate are independent of $M$. Moreover, we expect these two $\alpha$-dependent regimes to also exist in full spin foam models, albeit for different parameter values. Spin foam vertex amplitudes are (in the semi-classical limit) polynomially suppressed under uniform scaling of their representations \cite{Conrady-Freidel,FrankEPRL}, while the scaling behavior of the scalar field in terms of lengths is encoded in its action. Thus, while other effects will appear under lifting the cuboid restrictions, e.g. oscillatory spin foam amplitudes, the respective scaling behaviors persist, which depends on the combinatorics of the 2-complex.

\begin{figure}
    \centering
    \includegraphics[width = 0.45\textwidth]{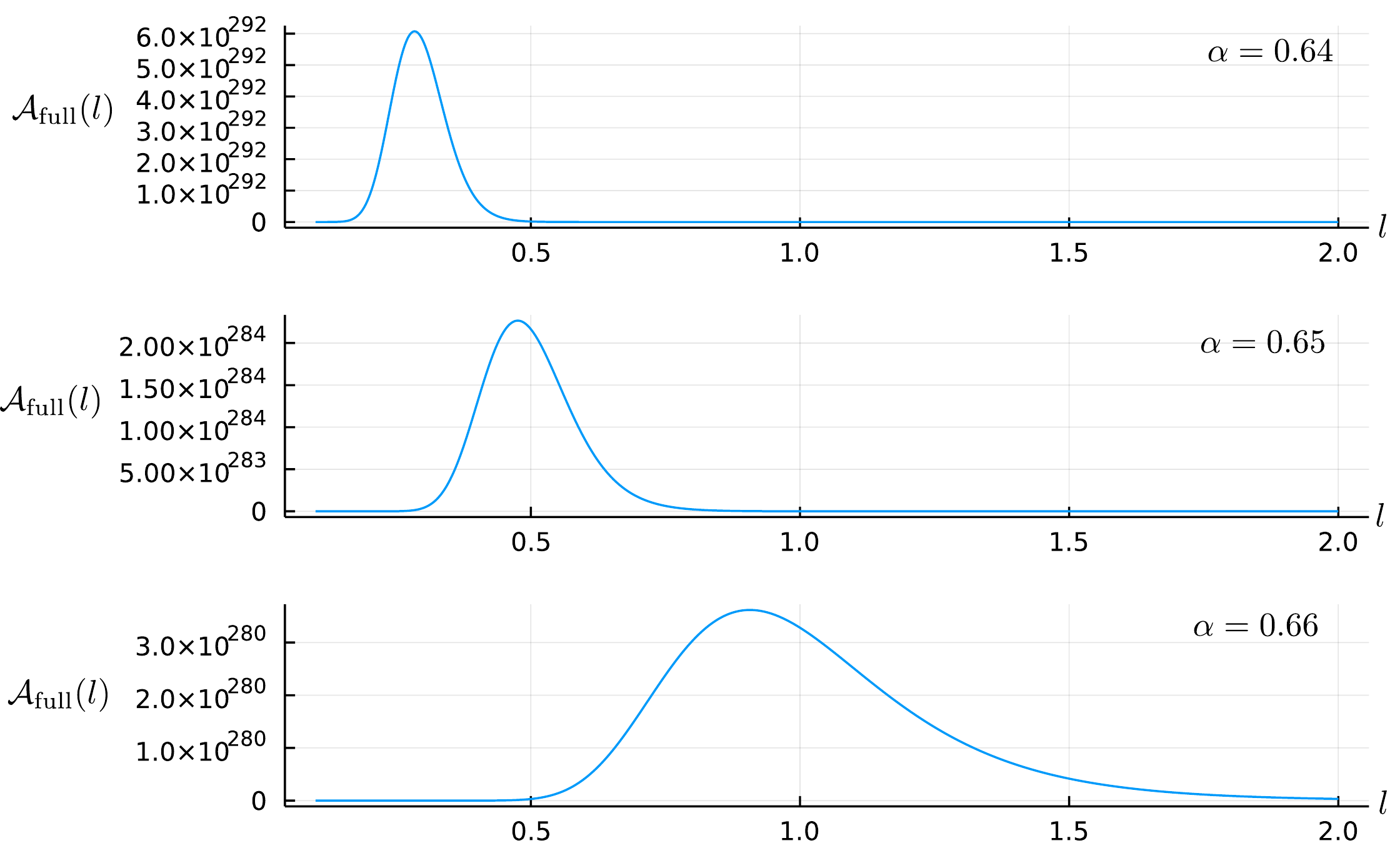} \; \;
    \includegraphics[width = 0.45\textwidth]{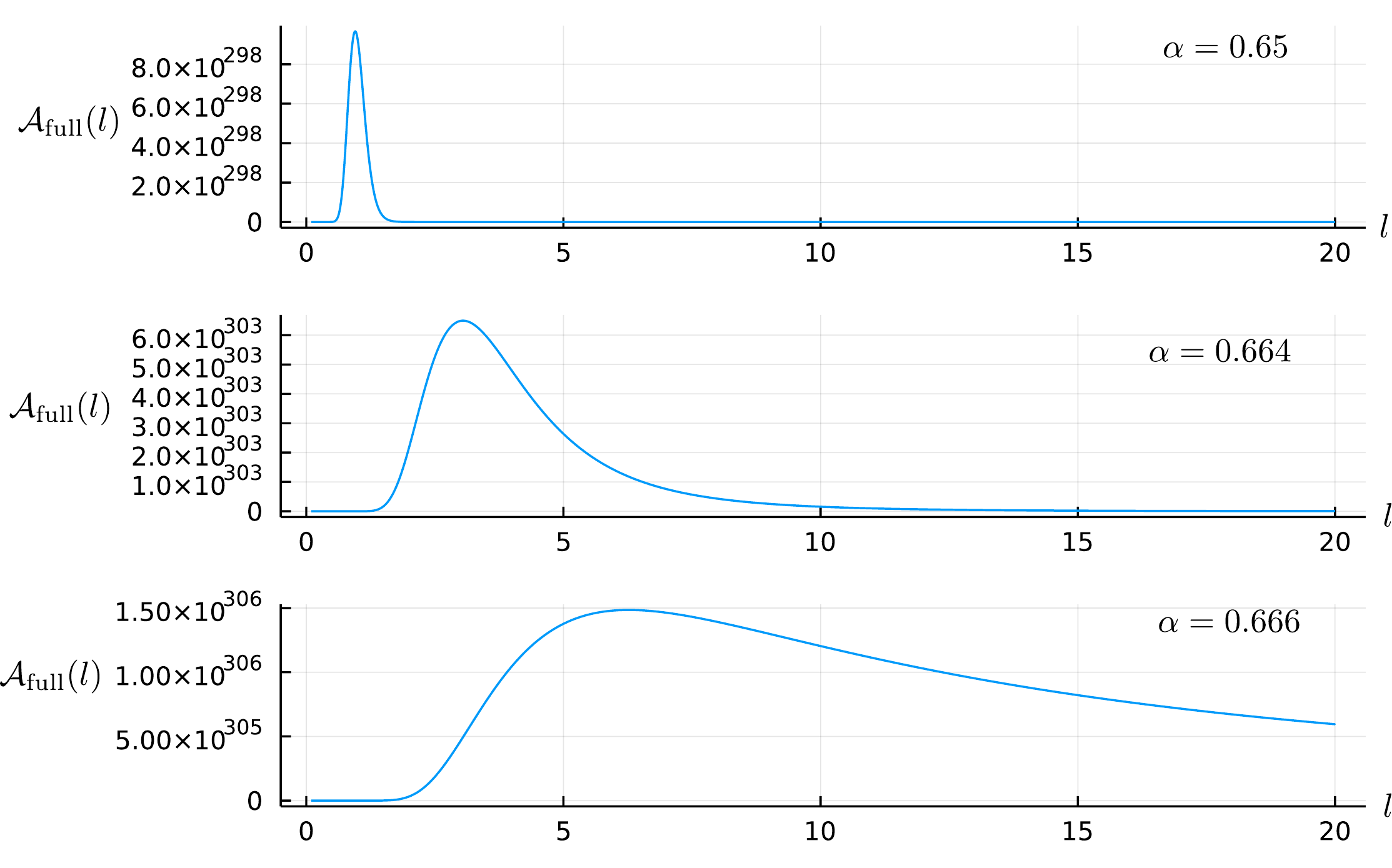}
    \caption{\textit{Left:} Plot of the full amplitude for $\alpha= 0.64, 0.65, 0.66$ well within the region in which the scalar field and spin foams balance each other out. The peak shifts to larger lengths as we increase $\alpha$ and the its width increases.
    \textit{Right:} As we further increase $\alpha$ the width of the peak starts to grow rapidly, here for $\alpha=0.65, 0.664, 0.666$. Both plots are for lattice size 3.}
    \label{fig:equal_lengths}
\end{figure}

The form of the denominator additionally suggests that an intermediate region between small and large $\alpha$ exists, where the length $l$ is finite. This is due to the various summands in the denominator: qualitatively, for small lengths the term with smallest exponent dominates while clearly the opposite is true for large lengths. If we choose a value for $\alpha$ such that the scaling behavior of the spin foam amplitude is between these two extremes, the probability distribution should have a peak at finite lengths, whose position and width is a function of $\alpha$ and the mass $M$. This is indeed the case as we show in fig. \ref{fig:equal_lengths}.

From the simple form of the probability distribution we can straightforwardly derive the interval in $\alpha$ for which we expect finite lengths. For small length $l$ the probability distribution scales like $l^{N (24 \alpha -15) - 1}$, whereas for large length it scales as $l^{N (24 \alpha -16) -1}$. Hence, for $N \gg 1$ the interval of interest is $\frac{5}{8} = 0.625 \leq \alpha \leq \frac{2}{3} = 0.66$. Of course, all of these considerations are strictly only true if all lengths are the same, but they provide us valuable guidance and insights into how the dynamics change once all lengths in the spin foam are dynamical, i.e. we consider a superposition of irregular lattices.

\subsection{Observables of the coupled system}
\label{Observables-of-interest}

The expectation values of observables are defined analogously to other path integrals or statistical systems. Given an observable $\mathcal{O}$, which depends on the configuration variables $\{l_i\}$ and/or $\{\phi_a\}$, its expectation values are straightforwardly defined as:
\begin{equation}
    \langle \mathcal{O} \rangle = \frac{1}{Z} \int \prod_i dl_i \, \int \prod_{a \in \sigma_0} d \phi_a \; \mathcal{O}(\{l_i\},\{\phi_a\}) \prod_{v \in \star \sigma_4} \hat{\mathcal{A}}_v \; e^{\sum_{a,b} \phi_a \, K_{ab} \, \phi_b} \quad .
\end{equation}
Here we have suppressed the dependence of the spin foam amplitudes and the matter action on the configuration variables to keep the expression readable. In this work we are considering observables of either the spin foam model or the scalar field. For the spin foam, we examine simple geometric observables like the total 4-volume $V$ or individual lengths $l_i$ and their variances respectively, while for the scalar field we are interested in the correlations $\phi_a \phi_b$ of the scalar field located at vertices $a$ and $b$. These observables are interesting for the following reasons.

The geometrical observables, like the total volume of space-time, are relevant as they are a direct indicator for the scaling behavior of the coupled model. Cuboid spin foams by themselves usually have diverging edge lengths / 4-volume for values of $\alpha$ above $\sim 0.58$; then the amplitude $\hat{\mathcal{A}}_v$ is a homogeneous function of positive degree and largest lengths dominate. However, we expect that the scalar field instead favours smaller edge lengths: the kinetic part of the action scales with $\sim l^2$, while the potential part with $\sim M^2 l^4$. Hence the action in the fields is smaller if the lengths are smaller, which gets more pronounced as the mass is increased. This interplay between the spin foam and matter part is intriguing.

On the other hand, a free, massive scalar field is solely determined by its 2-point correlation function; any correlations of higher degree can be understood as arising from 2-point correlations. Moreover, these correlations drop off exponentially with the distance between vertices and depend on the mass of the scalar field. Thus, it is the ideal observable to study here, yet, due to our setup, the interpretation is more subtle. Consider the correlator $\langle \phi_a \phi_b \rangle$ for fields located at vertices $a$ and $b$ in lattice field theory on a regular lattice. Here, the lattice inherits the notion of distances as a discretisation of the flat background space-time, such that the vertices $a$ and $b$ \textit{are} space-time events. We readily know the geodesic distance between these vertices and use this information to compute the correlation length from the correlations between all the vertices of the lattice. However, in a spin foam and more generally in a background independent setup, the vertices $a$ and $b$ have no inherent meaning; they are merely unphysical coordinates/labels. Their distance is not fixed by the discretisation, in fact we are integrating over superpositions of distances between these two vertices. Nevertheless, in addition to the correlations of the two fields located at $a$ and $b$, we also need the distance the vertices $a$ and $b$ are apart to obtain a physically meaningful relational observable \cite{Rovelli:2001bz,Dittrich:2005kc,Tambornino:2011vg}.

In our Monte Carlo simulations, we use the following approach: for each sample of lengths and fields, we compute the correlations of fields for all pairs of vertices and the distances between these vertices for a specific sample. Then we use the collection of data of correlations and distances to determine the correlation length as a non-local, coarse grained observable to learn how the matter theory effectively behaves on the spin foam background. Note that in this approach we cannot control the distance at which we probe the correlations; we can only access the correlations for \textit{probable} spin foam and scalar field configurations. A suggestion to measure the correlations at arbitrary geodesic distance is described in the introduction of \cite{Ambjorn:2012jv}\footnote{The idea is to define the correlator as a function of geodesic distance. Essentially one integrates the scalar field correlation function over all pairs of space-time points over all metrics and enforces a fixed geodesic distance by a delta distribution.}.

\subsection{Setup for numerical simulations} \label{sec:monte_carlo}

The partition function for the cuboid spin foam model with minimally coupled scalar degrees of freedom on the vertices of the cubulation is given by \eqref{partition-function}. In order to study this system we compute observables of the coupled system using Markov Chain Monte Carlo integration. The general idea of Monte Carlo integration is to approximate an expectation by the sample mean of the function of simulated random variables \cite{Metropolis_1949}. We are interested in computing the expectation values of observables of lengths and fields as defined in the previous section. We approximate this expectation value with the sample mean of $\mathcal{O}({l_i},{\phi_a})$, which is a function of the random variables (lengths and fields) drawn from the distribution 
\begin{equation}
\mathrm{P}(\{l_i\},\{\phi_a\}) = \frac{1}{Z}\prod_{v \in \star \sigma_4} \hat{\mathcal{A}}_v \; e^{\sum_{a,b} \phi_a K_{ab} \phi_b} \quad .
\end{equation}

We use the Metropolis-Hastings algorithm to sample the lengths and fields. This algorithm is based on a theorem that states that the Markov chain with transition probabilities arising from the Metropolis-Hastings algorithm converges to the target distribution if the chain is ergodic and satisfies detailed balance \cite{Hastings_1970}. In our setting each configuration of lengths and fields is an element of the Markov chain. As this convergence can take time we need to discard some initial samples (called the burn in) in order to ``forget'' the starting configuration.
Although the marginal distribution of the Markov chain converges to the target, it does not mean that the chain converges to a chain of identically independently distributed draws from the target distribution. This is because of the auto-correlation introduced by the proposal distribution leading to a bias in the observed sample means. Thus it is important to sample the chain at intervals longer than the auto-correlation length. This auto-correlation also introduces a trade-off between auto-correlation and acceptance rates. If the acceptance rate is too high the auto-correlation length increases as each consecutive sample represents only a small change in the target distribution. On the other hand a high rejection rate results in slower convergence to the target distribution.

For this coupled system it was particularly difficult to balance the acceptance-rejection ratio. Changing both the lengths and fields together leads to high rejection rates. This forced us to settle for smaller changes in the configuration to maintain a reasonable acceptance rate. We chose to either change all the lengths associated to a hypercuboid or the field at one vertex. New lengths were proposed by rescaling
\begin{equation}
    l_i^{new} = r_i l_i^{old}, \quad r_i \in \mathcal{U}(0.5,2)
\end{equation}
where the rescaling factor $r_i$ is different for each length of a hypercuboid. Choosing this rescaling factor between $0.5$ and $2.$ gives healthy acceptance rates ($40-60\%$) and also allows us to explore several orders of magnitude of edge lengths. 
Additionally we also restricted the configuration space by imposing an upper ($10^4$) and a lower cutoff ($5\times 10^{-5}$) on the lengths in order to avoid numerical overflow as the spinfoam amplitude diverges as the lengths go to $0$ or infinity \footnote{The physical meaning of these cut-offs is different. For the lower cut-off one might choose the minimal non-vanishing area eigenvalue of a face, however also vanishing areas are permitted in the state sum. Typically, the upper cut-off needs to be removed as it breaks gauge invariance, unless one is considering a model incorporating a cosmological constant, see e.g. \cite{turaev-viro,q-spinfoam2,q-spinfoam,Haggard:2015nat,Han:2021tzw}}. Detailed experimentation showed this did not have any qualitative impact on the results presented in this article.
New field values were proposed using
\begin{equation}
\phi_i^{new} = \phi_i^{old} + r, \quad r \in [-\delta, \delta]
\end{equation}
where $\delta$ was chosen to maintain an acceptance rate between $40-60\%$. The consequence of these incremental changes is high auto-correlation resulting in a longer burn-in period and the need for more intermediate Monte Carlo steps between samples. 

\begin{figure}
    \centering
    \includegraphics[width = 0.31 \textwidth]{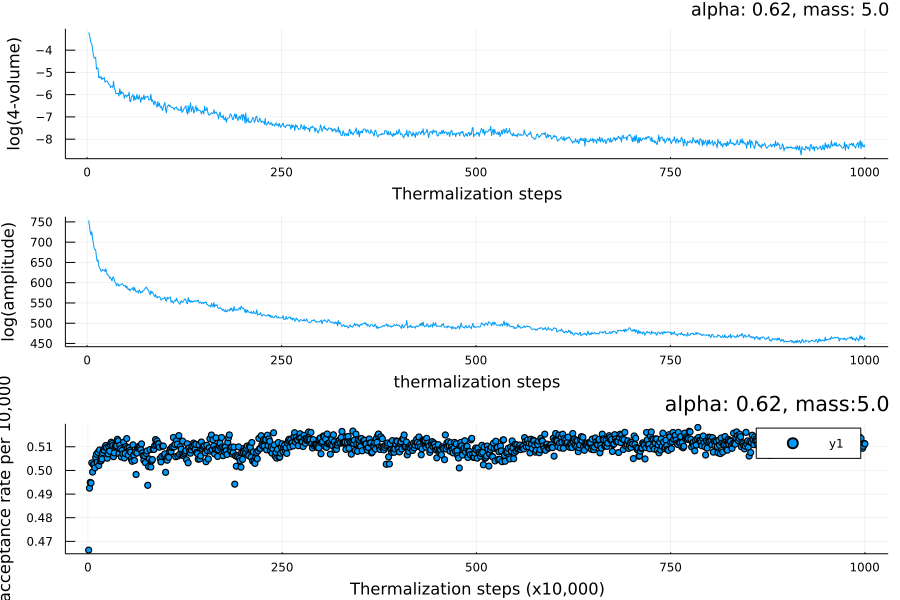} \; \;
    \includegraphics[width = 0.31 \textwidth]{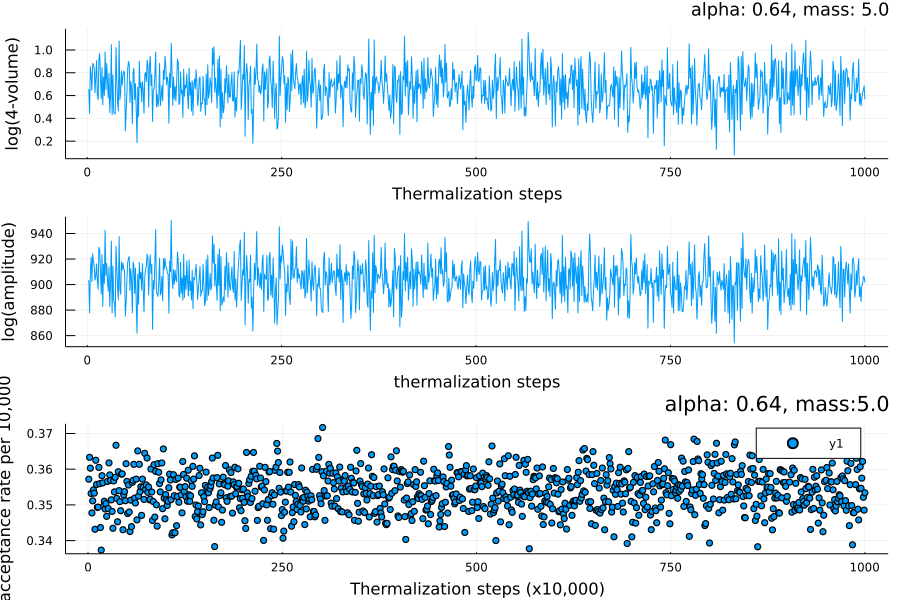} \; \;
    \includegraphics[width = 0.31 \textwidth]{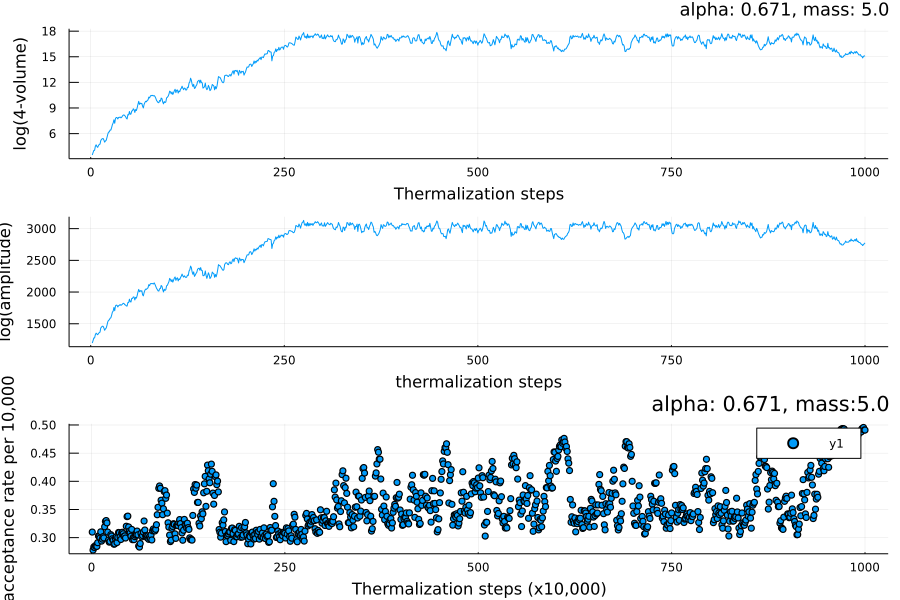}
    \caption{Plots of several functions during thermalization, logarithm of the total volume, logarithm of the total amplitude / integrand in the partition function and the average acceptance rate. Plots for lattice size $4$ and mass $M=5.$. \textit{Left:} $\alpha = 0.62$, outside plateau region and converges to lower cut-off. \textit{Centre:} $\alpha =0.64$, inside plateau region, system appears to have thermalized. \textit{Right:} $\alpha = 0.67$, outside plateau region, converges to upper cut-off.}
    \label{fig:thermalization}
\end{figure}

In order to determine if the chain had thermalized we plotted the amplitude and the volume for each Monte Carlo step in fig. \ref{fig:thermalization} for different distinctive regions of the model. During the burn-in period we expect a trend (increasing or decreasing) in the amplitude plots but after thermalization we expect the amplitude to be uniformly distributed over some range.

\begin{figure}[h!]
    \centering
    \includegraphics[width=0.95\textwidth]{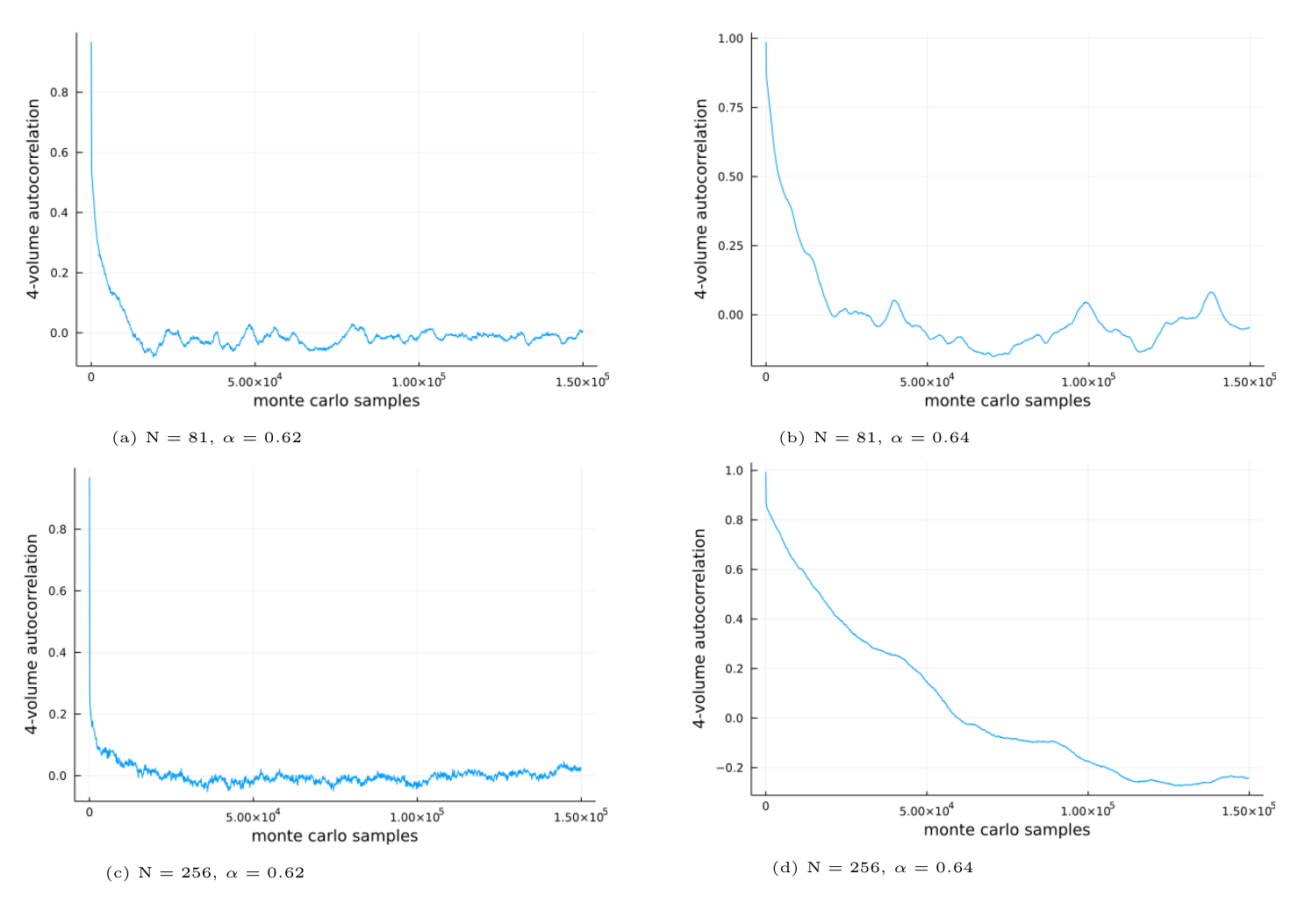}
    \caption{Plots of the total $4$-volume autocorrelation for different cubulation sizes and $\alpha$ values. For $\alpha$ values outside the plateau region where all the lengths approach the lower or upper cutoff, the autocorrelation for all lattice size is low. Within the $\alpha$ range where the lengths remain bounded the autocorrelation quickly increases with the number of cubulation vertices ($\textrm{N}$) as seen by the figures in the right column.}
    \label{fig:autocorrelationplot}
\end{figure}

After thermalization, expectation values of observables can be computed by drawing samples from the Markov Chain at regular intervals (determined by the auto-correlation length). In fig. \ref{fig:autocorrelationplot} we plot the auto-correlation over the samples for different parameters. There are four main sources of error in this estimate:
\begin{itemize}
\item Estimating the state of thermalization of the chain incorrectly.
\item Highly correlated samples
\item The Markov chain takes too long to explore some parts of the distribution and thus these states are not well represented in the sample (the ``missing mass" problem).
\item Not sampling enough
\end{itemize}
The first two errors are easily controlled for. The first by identifying suitable observables to test for thermalization. We use the full amplitude and the total volume of the spin foam for this. 
The second error can be mitigated by sampling the Markov Chain at intervals greater than at least twice the autocorrelation length for the observable of interest, where the autocorrelation length is computed after thermalization. This is known as sub-sampling.

The ``missing mass" problem is harder to identify and control for. A dirty fix is to run multiple simulations each time starting with a different configuration. We tested our algorithm by running multiple simulations for randomly chosen initial configurations for lattices with $81$ vertices. But the time cost was prohibitive for larger lattices. 

The last problem stems from the fact that the error in the estimate of the expectation value of the observable is dependent on the number of samples. Since sub-sampling minimizes the auto-correlation between consecutive samples drawn from the Markov Chain, we can treat these samples as independent. Then the true expectation value of the observable of interest in terms of the sample mean of a sample of size $N$ can be written as:
\begin{equation}
    \langle O \rangle = \frac{1}{N}\sum_{n=1}^{n=N} O_n + \frac{\sigma^2(O)}{N^2}
\end{equation}
where $O_n$ is the observable value calculated using the $n^{th}$ sample. The second term is the error term in our estimate of $\langle O \rangle$, where $\sigma$ is the standard deviation of $O$. Thus the error term can be reduced simply by increasing the number of samples. We must thus balance the time cost of generating the samples with the error in estimating the expectation value by the sample mean.

\begin{figure}
    \centering
    \includegraphics[scale=0.65]{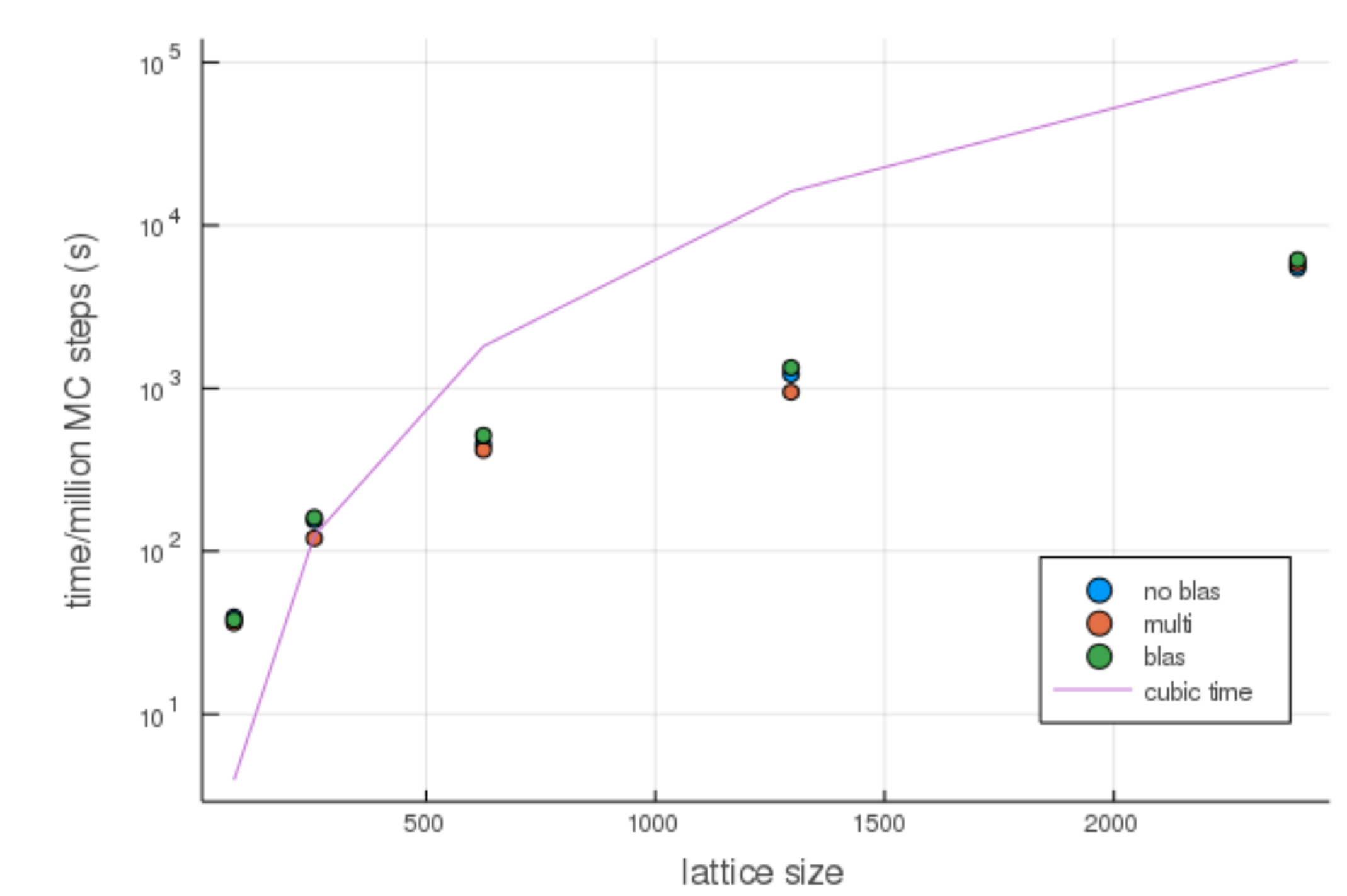}
    \caption{Plot of the time taken to simulate $1$ million Monte Carlo steps with respect to the number of vertices. The line denotes $O(N^3)$ time complexity. As expected the actual time complexity is $< O(N^3)$ even for the single threaded (no BLAS) implementation. }
    \label{fig:benchmarking}
\end{figure}

The time complexity of the algorithm scales linearly with the number of Monte Carlo steps and as a polynomial of the number of vertices. Each time we sample the lengths we need to recalculate the amplitude of the coupled system. This is the most expensive operation as calculating the spin foam amplitude is $O(N)$ for a cubulation of $N$ vertices and the calculation of the scalar field action involves a matrix multiplication which in the worst case results in $O(N^3)$ and even in the best case scales much faster than $N^2$. We optimize the algorithm to some extent by directly computing the change in the scalar field action when only the fields are sampled instead of computing the full action. However, this does not impact the worst case time complexity which remains $O(MN^3)$ where $M$ denotes the number of Monte Carlo steps and $N$ is the number of vertices in the cubulation. In actuality the performance is not quite so dire because the matrix multiplication algorithms are highly optimized and multithreaded while the analysis above only applies to calculations on a single thread, see fig. \ref{fig:benchmarking}. Even so it elucidates the difficulty in exploring large cubulations for which thermalization is slower (more Monte Carlo steps) and auto-correlations persist over longer ranges, see fig. \ref{fig:autocorrelationplot}, resulting in a higher number of Monte Carlo steps between each sampling step.

When the observable is only a function of the lengths it is possible to sum over the scalar fields analytically and then apply the Metropolis algorithm to evaluate the sum over the lengths. Sampling over just the lengths is easier as the burn-in period is shorter and the Markov chain can be sampled more frequently since the auto-correlation is lower. Details are provided in the next section.

For this system, the observables we are interested in are detailed in Section \ref{Observables-of-interest}. The geometric observables can be calculated both by sampling the lengths and fields as well as by analytically summing over the fields and then sampling the lengths. We compute the $4$-volume of the spinfoam to investigate the interplay between the scalar field amplitude which favours small lengths and the cuboid spinfoam amplitudes, which favours large lengths as $\alpha$ is increased. We find that there exists a range of $\alpha$ where the opposing effects of the scalar field amplitude and the spinfoam amplitude balance resulting in finite values of the $4$-volume. These results are detailed in Section \ref{section:geometricObservables}.

The most important observable of the fields is the 2-point correlation function. We compute the correlation function $\langle \phi_a\phi_b \rangle$ by averaging the product of the field values over all pairs of vertices $a$ and $b$. We find that when plotted as a function of the geodesic distance between vertices $a$ and $b$ this correlation function falls off exponentially with distance, much like the $2$-point correlation function on a fixed lattice, for the range of $\alpha$ where the $4$-volume remains finite.

\subsection{An alternative algorithm}

In addition to the algorithm described above, in which both lengths and fields are treated by Monte Carlo techniques, an alternative algorithm is possible. Independent of the shape of the underlying spin foam, the action of the scalar field is always of the form $\phi_a \, K_{ab} \, \phi_b$. Thus, if the length are fixed, the integrals over the fields are Gaussian and can be straightforwardly performed. Hence, we can rewrite the partition of the coupled system purely as an integral over the lengths:
\begin{equation}
    Z = \int \prod_i dl_i \, \frac{1}{\sqrt{\det K}}  \prod_{v \in \star \sigma_4} \hat{\mathcal{A}}_v(\{l_i\}_{i \in \sigma_4})  \quad .
\end{equation}
In this representation, we would use $Z^{-1} \frac{1}{\sqrt{\det K}}  \prod_{v \in \star \sigma_4} \hat{\mathcal{A}}_v(\{l_i\}_{i \in \sigma_4})$ as the probability distribution to sample the lengths. The advantage is apparent: this distribution includes the full dynamics of the scalar field and thus its impact on the gravitational theory. This is particularly helpful for parameter ranges for which the scalar field is difficult to sample, namely small and large $\alpha$. Moreover, sampling just over the lengths is less sensitive to how new length configurations are proposed, auto-correlations decay more rapidly and the burn-in period is shorter as well. 
Therefore, this algorithm is well suited to study geometric observables, such as the expectation values of lengths or the total volume.

However, computing observables of the field is more involved. Of course, the 2-point partition function for fixed length is simply given by the inverse matrix of $K$, such that we obtain for its expectation value:
\begin{equation}
    \langle \phi_a \phi_b \rangle = \frac{1}{Z} \int \prod_i dl_i \, K^{-1}_{ab}  \prod_{v \in \star \sigma_4} \hat{\mathcal{A}}_v(\{l_i\}_{i \in \sigma_4}) \quad .
\end{equation}
This expression is clearly different than for geometric observables, since the factor $\sqrt{\det K}^{-1}$ is missing. To sample the lengths for this observable, we could use $Z^{-1} \prod_{v \in \star \sigma_4} \hat{\mathcal{A}}_v(\{l_i\}_{i \in \sigma_4})$. However, this is actually not a probability distribution, since the $Z$ is the partition function of both matter and spin foams. Moreover, we would actually only use the spin foam amplitude to sample length without considering the effect of the scalar field, which leads to dramatically different results.

Alternatively, we can attempt to use the previous probability distribution by expanding the integrand with $\frac{\sqrt{\det K}}{\sqrt{\det K}} = 1$; the denominator is part of the probability distribution, while we absorb the numerator into the observable. Unfortunately, this idea is not numerically feasible. Generically $\det K$ is a large quantity, even for relatively small lattice sizes. Hence, it completely dominates the expectation value and convergence of the results is slow. Therefore, we refrain from using this representation for Monte Carlo simulations.

\section{Results} \label{sec:results}

The simulations were performed at the Ara-Cluster at FSU Jena. For each set of parameters ($\alpha$, M and $N$), we randomly generated an initial configuration of fields and lengths followed up by a burn-in of $10^8$ Monte Carlo steps. This was more than sufficient for the system to thermalize. Then, depending on the lattice size we sampled the Markov chain at varying intervals to ensure that the samples were not correlated. We chose the following interval lengths by explicitly measuring the autocorrelation lengths, i.e. the number of steps until two consecutive samples are uncorrelated and can be considered independent:
\begin{itemize}
    \item $N=81$: $5 \cdot 10^4$ Monte Carlo steps,
    \item $N=256$: $10^5$ Monte Carlo steps,
    \item $N=625$: $2 \cdot 10^5$ Monte Carlo steps.
\end{itemize} 
Due to the increase in auto-correlation length with lattice size and the increase in the time per iteration of the algorithm, we reduced the number of samples taken for lattice size $N=625$ to $2000$ while $5000$ samples were generated for  lattice sizes $N=81$ and $N=256$.

The algorithm and the data generated for and presented in this article are freely available. The algorithm can be found on Github\footnote{\url{https://github.com/amoosam/CuboidSpinfoamScalarField}}, whereas the data is available on Zenodo\footnote{\url{https://doi.org/10.5281/zenodo.6838757}}.

\subsection{Geometric observables} \label{section:geometricObservables}
\begin{figure}[h!]
    \centering
    \includegraphics[width=0.45\textwidth]{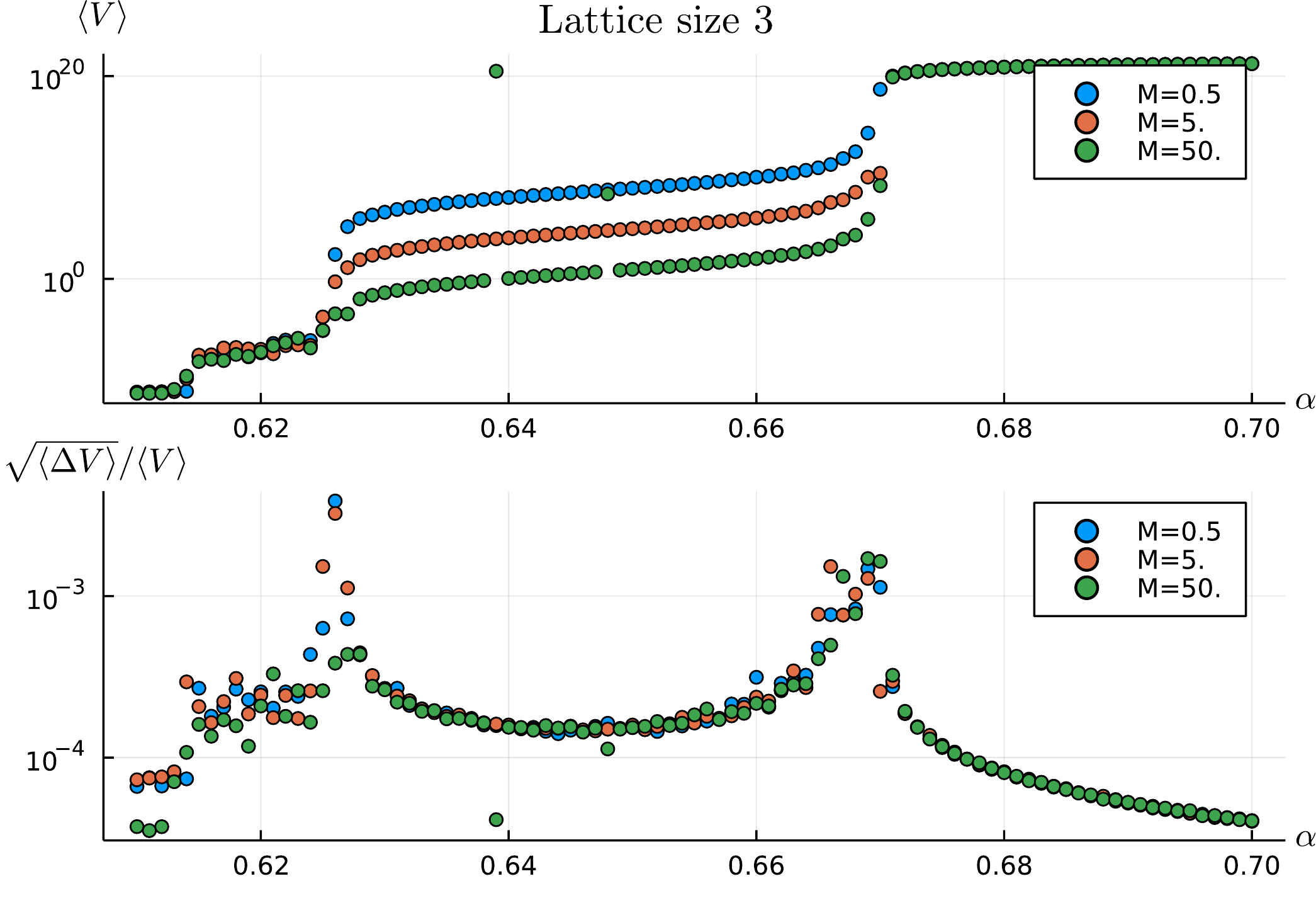} \; \;
    \includegraphics[width=0.45\textwidth]{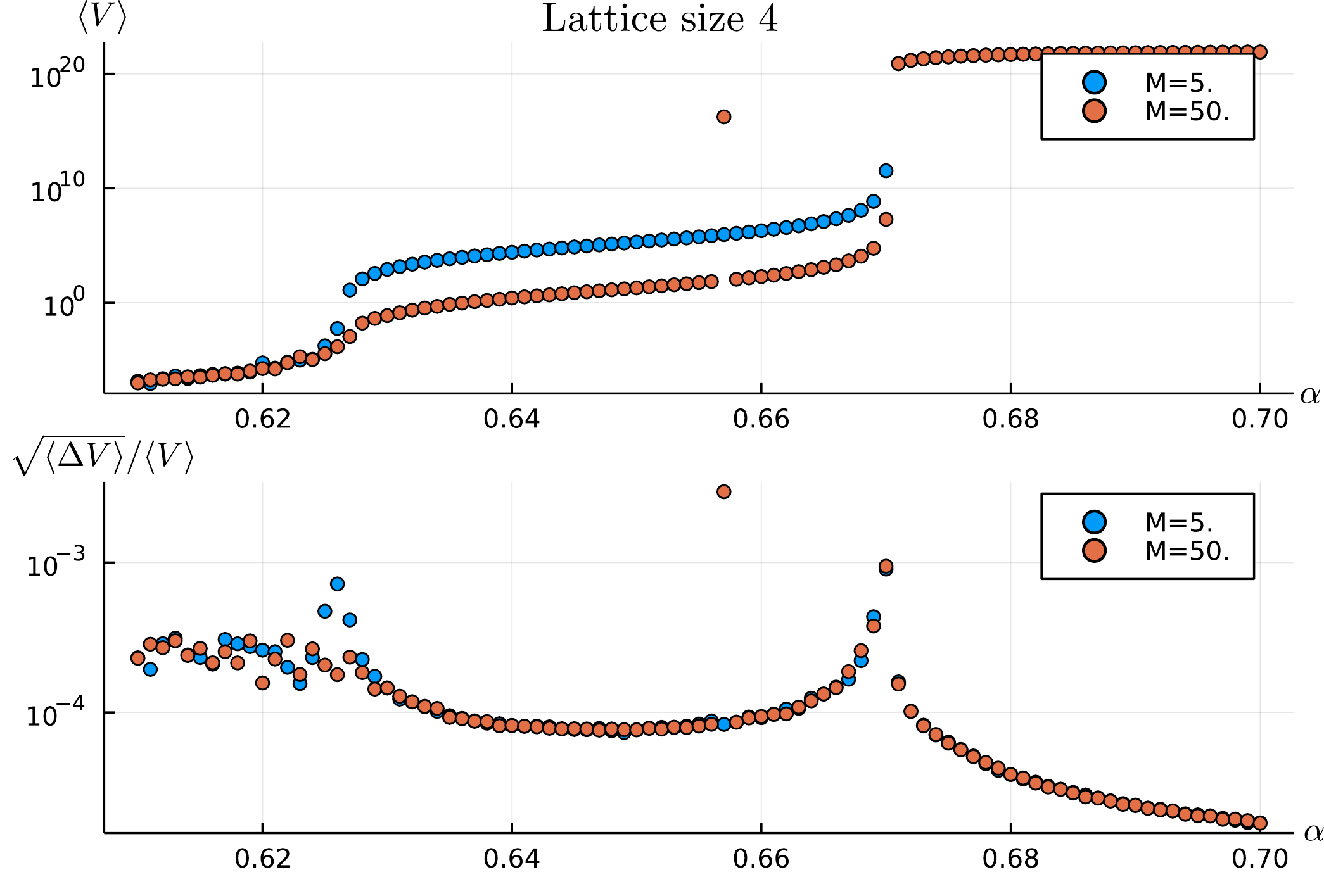}
    \caption{Plots of the expectation values of the total volume and its variance (normalized by the expectation value of the volume) as a function of $\alpha$ for different masses. \textit{Left:} Lattice size 3 ($N=81$). \textit{Right:} Lattice size 4 ($N=256$).}
    \label{fig:vol_plots}
\end{figure}

\begin{figure}
    \centering
    \includegraphics[width=0.45\textwidth]{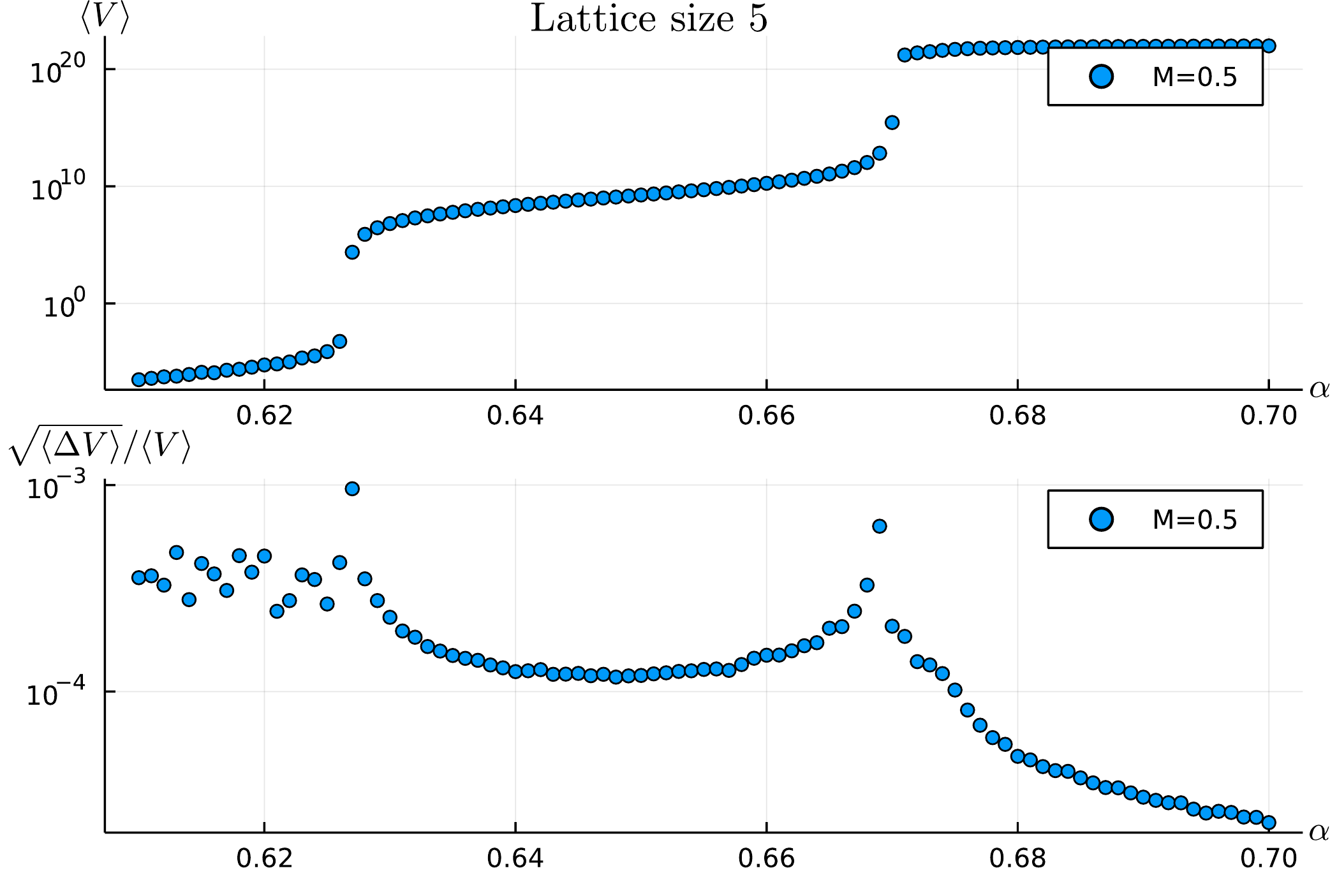}
    \caption{Plots for expectation values of total volume and its variance for $M=0.5$ and lattice size $5$ ($N=625$).}
    \label{fig:vol_plot_5}
\end{figure}

In section \ref{sec:qualitative_results} we obtained some qualitative insights for the scenario when all lengths are constrained to be equal. To check whether these qualitative insights also apply to the full model, expectation values of coarse geometric observables provide a good point of comparison. A suitable observable is the total four dimensional volume of the universe and the square root of its variance (normalized with respect to the volume expectation value). In the figs. \ref{fig:vol_plots} and \ref{fig:vol_plot_5} we plot both expectation values as a function of the parameter $\alpha$ for lattices of size $N=81$, $N=256$ and $N=625$ for different values of the scalar field mass $M$.

We observe the same qualitative behavior in all cases, independent of the mass and lattice size. For $\alpha < 0.63$ the total volume is small, close to the volume given by the minimal value of the lengths prescribed by the lower cut-off. On the other hand, for $\alpha > 0.67$ the volume is large and close to the volume for largest allowed edge lengths (the upper cut-off). These two regions are clearly cut-off dependent, such that our simulations do not reflect the actual dynamics in these regions. But they suggest that for small $\alpha$ the lattice lengths are as small as possible, while they are as large as possible for large $\alpha$ consistent with our previous observations.

The interesting regime lies between these two extremes: 
the boundary of the intermediate region is marked by steep increases of the average total volume, from the lower cut-off to a finite volume around $\alpha \approx 0.63$ and from a finite volume to the upper cut-off around $\alpha \approx 0.67$. The finite average volume is a function of $\alpha$ and the scalar field mass $M$. Increasing $\alpha$ gives a larger volume, while a larger mass drastically impacts the average volume.
Crucially, this region is independent of the choice of upper and lower cut-offs on the lengths and thus accurately captures the dynamics of the system. 
Due to the high symmetry of the cuboid model, we thus expect it to look on average like a regular lattice with an emergent lattice scale. To confirm whether this average is a good representative of the geometry, further observables must be examined. Note that these results are in good agreement with our previous considerations of the equilateral system in section \ref{sec:qualitative_results}. Moreover, the cut-off independent existence of a regime of finite volume is of note, as this does not exist for cuboid spin foams without matter \cite{Bahr:2015gxa}.

To learn more about the typical geometry of the spin foam, if such a notion makes sense, we also examine the (square root of the) variance of the volume presents an interesting feature: for smallest and largest volumes as well as the finite plateau region, the variance of the total volume is small. This has different meaning in the different regions: when the total volume is determined by upper or lower cut-off, the variance is small due to the cut-off. E.g. at the lower cut-off, the probability distribution would favor even smaller lengths, however such Monte Carlo steps are forbidden by construction. On the other hand, larger lengths are highly improbable. In contrast, the interpretation in the finite volume region is different: there all length variations are a priori permitted, yet large deviations away from the average configuration are not probable. Thus, we can conclude that the spin foam is peaked upon an emergent regular lattice with small deviations around it.

However, at the boundaries of the plateau, where the system transitions from a finite to a ``cut-off'' volume, the variance shows pronounced peaks. 
This is because the system transitions from being peaked at a finite volume to being peaked on the lower / upper cut-off. Therefore, the volume is permitted to vary in some cases over several orders of magnitude. The height and position of the peaks might be cut-off dependent, as we limit the minimal / maximal volume the system is allowed to take. This regime is interesting, as these peaks might indicate a phase transition of the system with the possibility of taking a continuum limit. More investigations are necessary though, e.g. considering larger systems.

\subsubsection{Comparing different lattice sizes}

\begin{figure}[h!]
    \centering
    \includegraphics[width = 0.45 \textwidth]{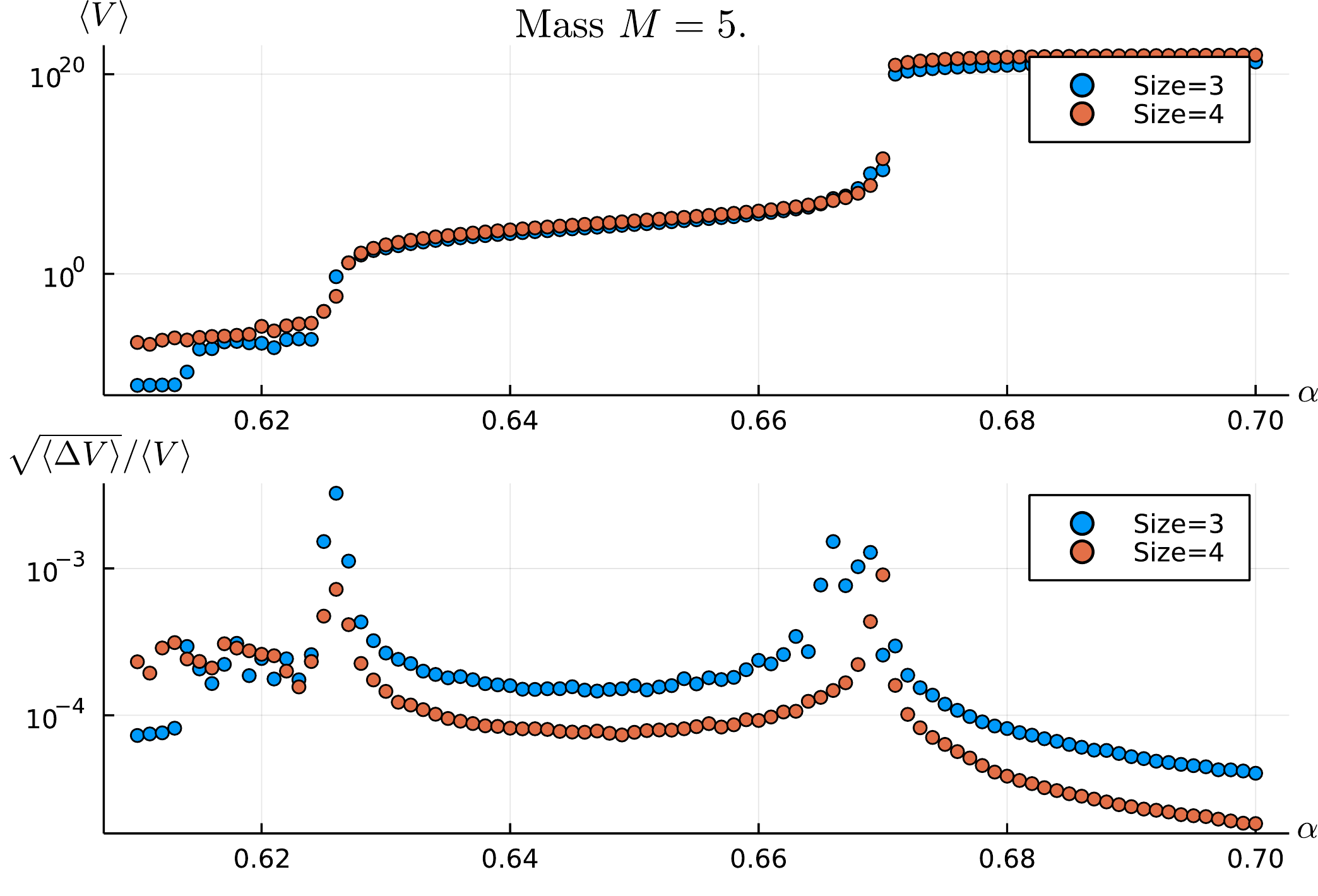} \; \;
    \includegraphics[width = 0.45 \textwidth]{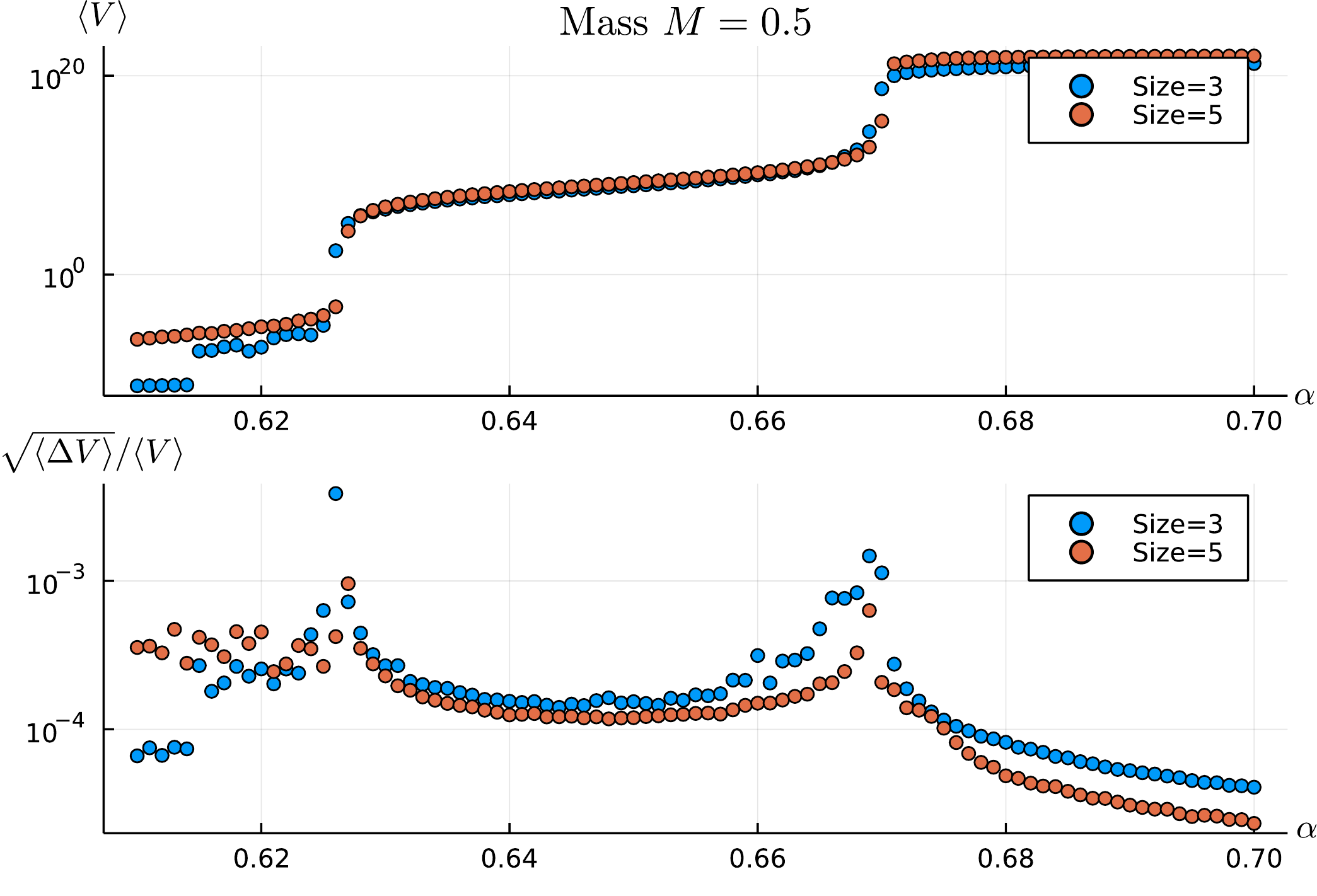}
    \caption{Comparison of volume and its variance for the same mass and different lattice sizes. \textit{Left:} Mass $M = 5.$ and comparison of observables for lattices with $3$ and $4$ independent edge lengths, i.e, $N=81$ and $N=256$. \textit{Right:} Mass $M=0.5$ and comparison of observables for lattices with $3$ and $5$ independent edge lengths, i.e. $N=81$ and $N=625$.
    }
    \label{fig:comparison_vol}
\end{figure}

In addition to the previous plots where we studied the total volume and its variance for different masses and the same lattice size, we briefly compare these quantities for the same masses and different lattice sizes. In fig. \ref{fig:comparison_vol} we compare lattice size $N=81$ ($3$ independent edge lengths) and $N=256$ ($4$ independent edge lengths) for a mass $M=5.$ and lattice $N=81$ and $N=625$ ($5$ independent edge lengths) for a mass of $M=0.5$.

Qualitatively, there is little difference between different lattice sizes. In the plateau region, the larger lattice size typically has a larger total volume, which indicates that the typical lattice lengths are similar for the same parameters. Moreover, the beginning and end of the plateau region is slightly shifted towards smaller $\alpha$ for larger lattices. This matches our expectations from our regular lattice considerations in section \ref{sec:qualitative_results}. We can confirm these observations also by the position of the peaks in the variance plots.

\subsubsection{Distribution of edge lengths}

From the finite expectation value of the total volume in the plateau region it immediately follows that the spin foam edge lengths are finite as well. Moreover, the low variance (compared to the volume expectation value) implies that the deviations are rather small. These results are compatible with our analytic considerations in section \ref{sec:qualitative_results}, where we fixed all edge lengths to be the same. Due to the high degree of symmetry of the system a finite total volume suggests that effectively the coupled system behaves like a regular lattice, whose lattice length is emergent and determined by the parameters of the model ($\alpha$ and $M$), plus small deviations. However, the total volume is a coarse observable and might average out other features. To complement our understanding of the system, we consider the distribution of lengths.

\begin{figure}
    \centering
    \includegraphics[width=0.45\textwidth]{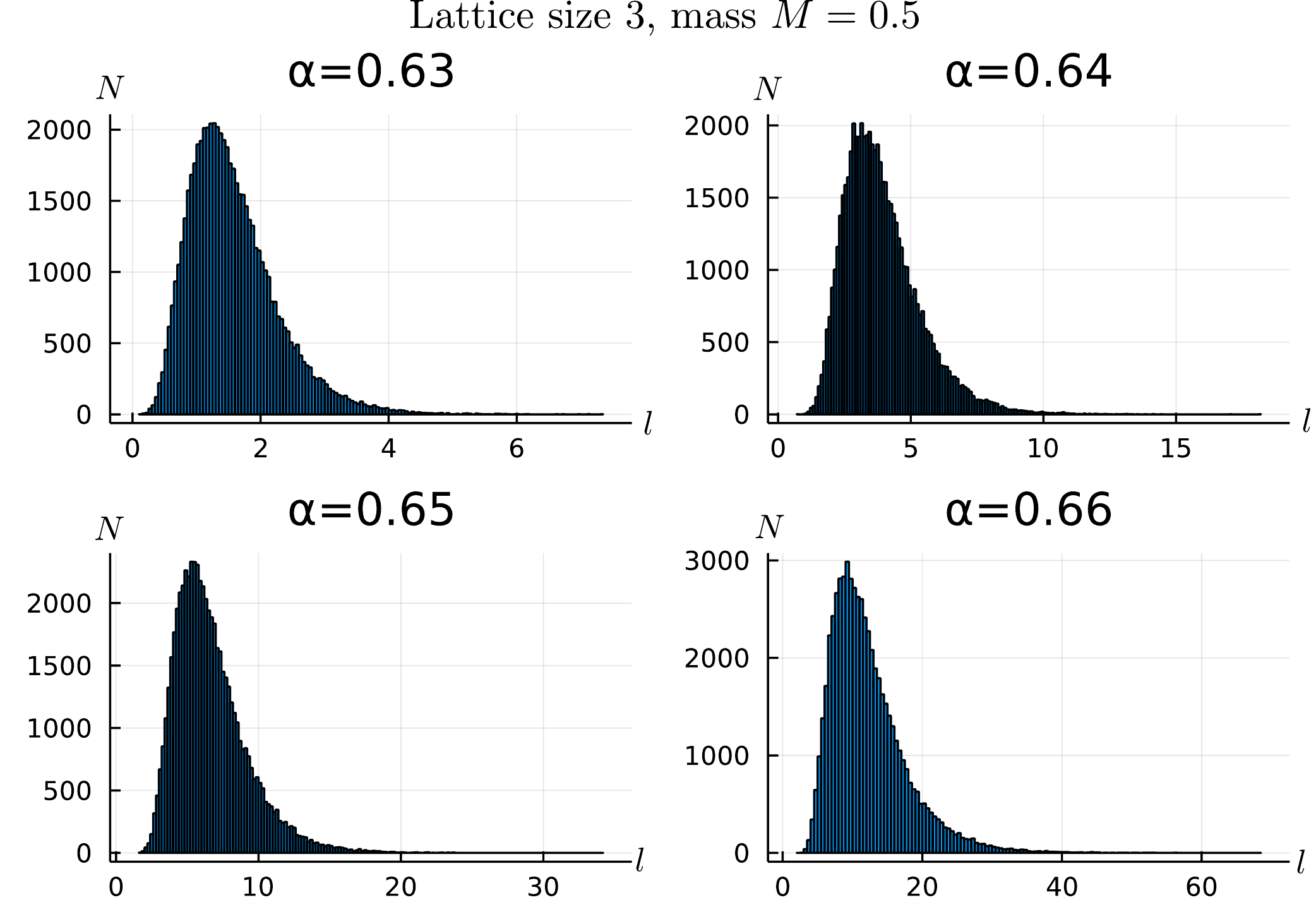} \; \;
    \includegraphics[width=0.45\textwidth]{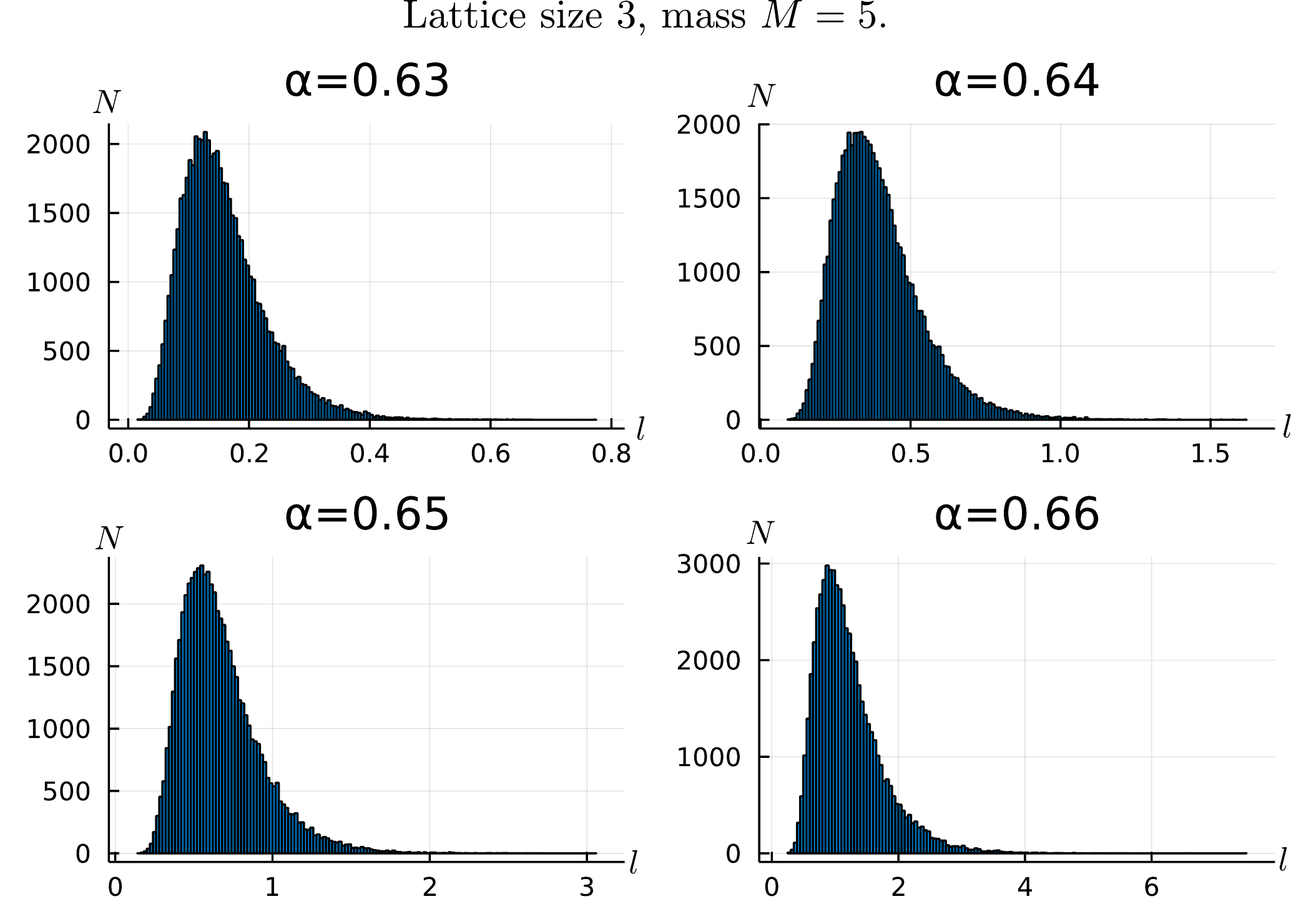}
    \caption{Histograms of all lengths (lattice size $N=81$) obtained by Monte Carlo simulations. \textit{Left:} Mass $M=0.5$. \textit{Right:} Mass $M=5.$}
    \label{fig:length_histogram_3}
\end{figure}

\begin{figure}
    \centering
    \includegraphics[width=0.45\textwidth]{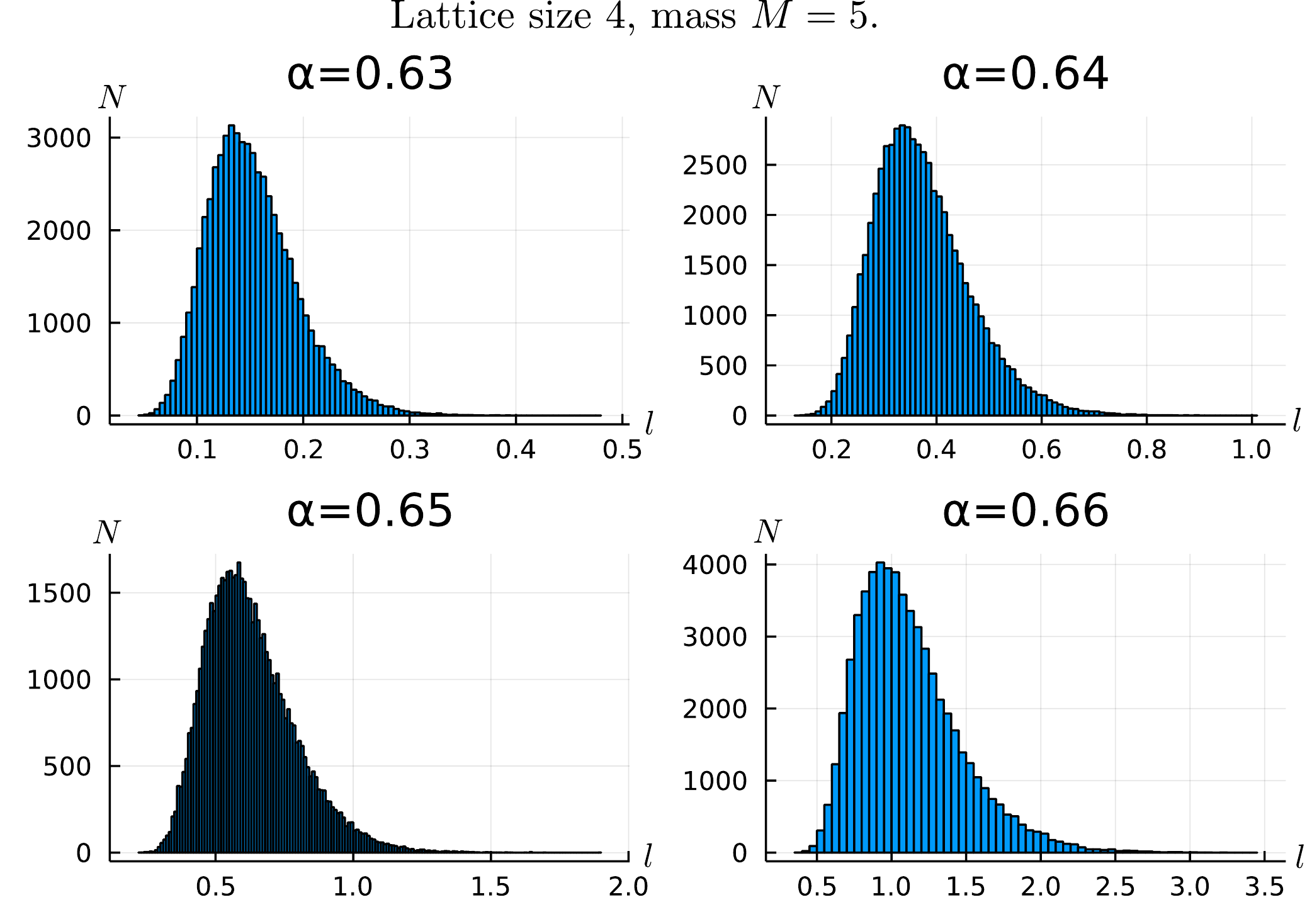} \; \;
    \includegraphics[width=0.45\textwidth]{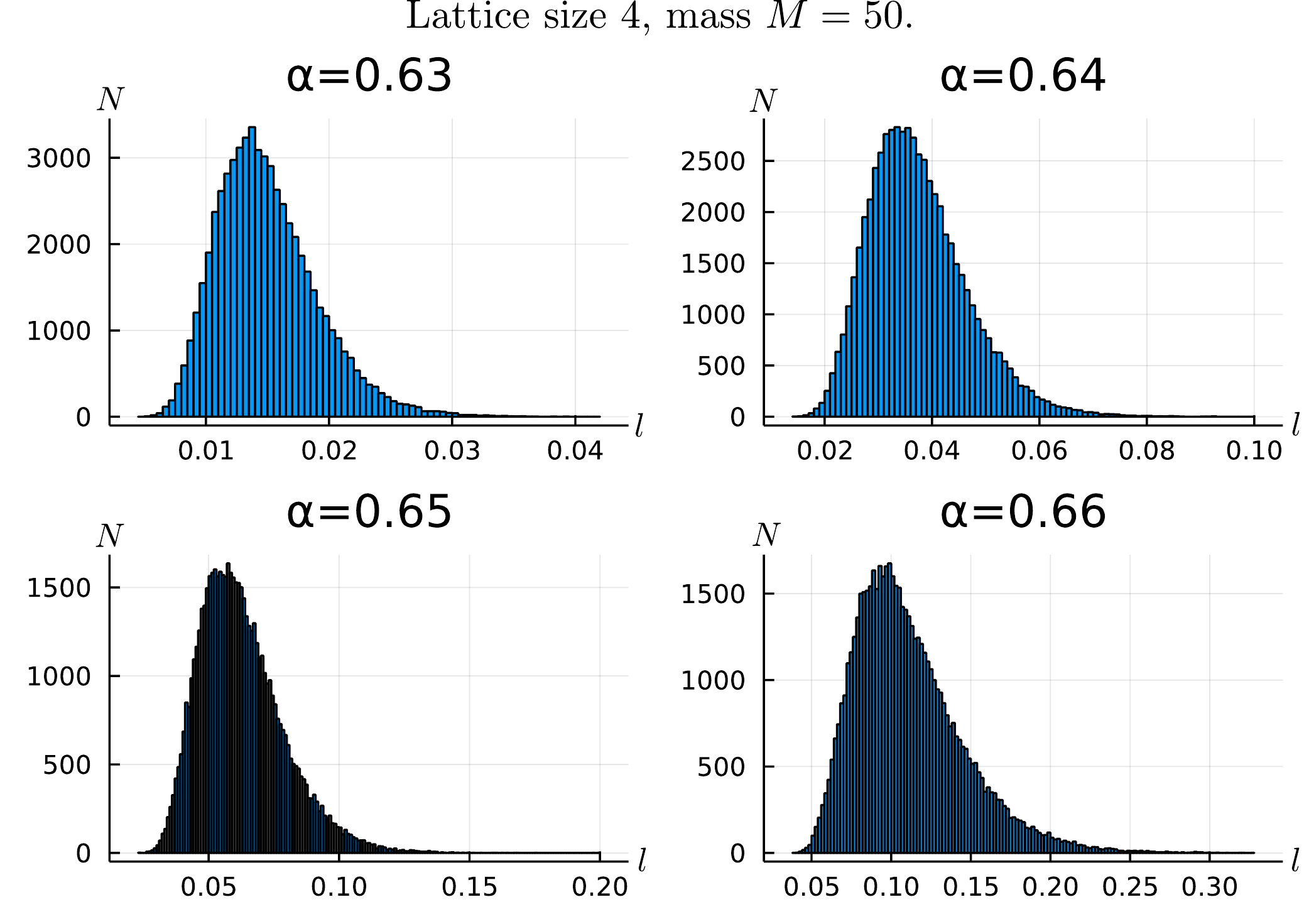}
    \caption{Histograms of all lengths (lattice size $N=256$) obtained by Monte Carlo simulations. \textit{Left:} Mass $M=5.$. \textit{Right:} Mass $M=50.$}
    \label{fig:length_histogram_4}
\end{figure}

\begin{figure}
    \centering
    \includegraphics[width=0.45\textwidth]{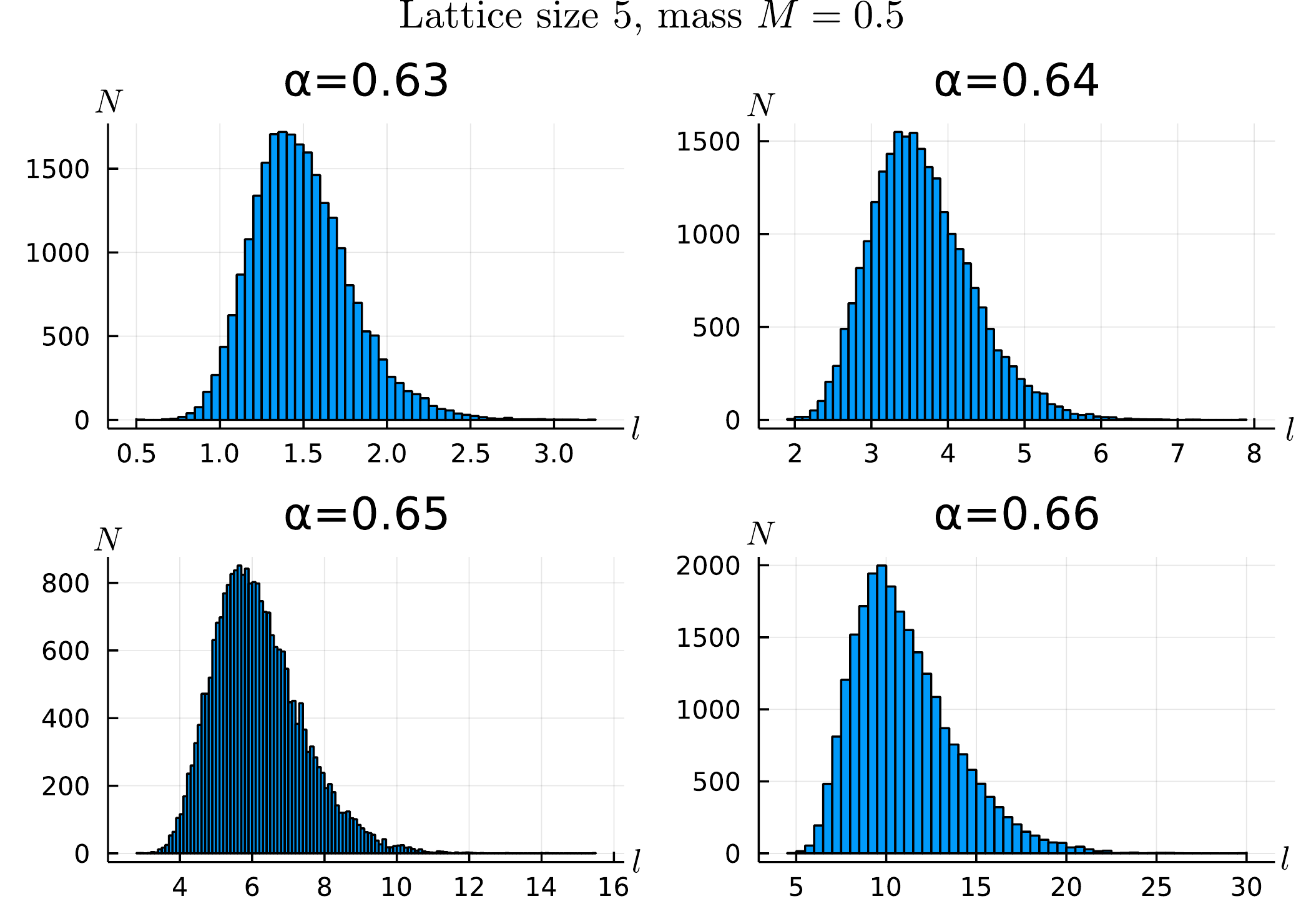}
    \caption{Histograms of all lengths (lattice size $N=625$) obtained by Monte Carlo simulations.}
    \label{fig:length_histogram_5}
\end{figure}

To do so, we plot histograms of the length distribution for different values of $\alpha$ (in or close to the plateau region), $M$ and the lattice size in figs. \ref{fig:length_histogram_3}, \ref{fig:length_histogram_4} and \ref{fig:length_histogram_5}. In these histogram we show how often all lengths, i.e. in all space-time directions, were sampled in an interval. From a qualitative perspective, the plots are similar for any lattice size, mass $M$ and $\alpha$: we see a (fairly) smooth distribution with a clearly defined peak at a finite length; some plots look more jagged, which is due to the finite sample size (in particular for lattice size $5$). The peaks are asymmetric, with a sharper decline to shorter lengths and a long tail towards larger lengths in good agreement with our analytical examinations in section \ref{sec:qualitative_results}. In short, this distribution shows that in the plateau region all lengths are sharply peaked around the same value, which confirms our expectation that the system is peaked on an emergent regular lattice. Only small deviations are permitted, such that we are confident to characterize the effective spin foam geometry as a regular lattice plus perturbations. In the following, we will discuss how this effective geometry changes as we change the parameters of the model.

While all plots look fairly similar, there are strong quantitative differences when changing the parameters. Most striking are changes in the mass $M$: increasing it by an order of magnitude moves the peak of the length distribution one order of magnitude lower for all lattice sizes and values of $\alpha$ (in the plateau region), and suggests a linear relation between the mass and the position of the peak. However, while the absolute variance of lengths decreases, relative to the peak of the distribution the variance appears to be the same. In this sense, a larger mass ``shrinks'' the distribution inversely proportional to the mass increase.

In contrast, increasing the value of $\alpha$, while keeping the mass and lattice size fixed, moves the peak of the length distribution to larger lengths. Moreover, while we always see a sharp peak in the plateau region, the tail towards large lengths becomes longer as we increase $\alpha$ suggesting a larger variance. This is also confirmed by the variance of the total volume. However, compared to our analytical study of superimposed regular lattices, the tail appears to be slightly shorter. However this impression could be misleading, since these lengths are less probable and might require larger sample sizes to properly reflect the size of these tails.

Finally, we consider the changes of the lengths distribution for different lattice sizes. Firstly, if we keep the mass $M$ and $\alpha$ fixed, the peak of the distribution appears to be more or less constant, that is the typical edge length is mostly determined by $M$ and $\alpha$ and not by the lattice size. Clearly, it is necessary to go to larger lattice sizes to confirm that. Additionally, it seems that the distribution becomes more sharply peaked for larger lattice sizes, in particular the tail is shorter. Again, though more samples might be necessary to properly represent the distribution at such low probability, there is another reason why larger lattice sizes can be more restrictive.

Due to the high degree of symmetry of quantum cuboids, the entire spin foam is characterized by $4 N$ edge lengths, where $N$ is the lattice size. If we vary a single edge length in the spin foam, we affect $N^3$ hypercuboids and many scalar field interactions via the change of edge lengths directly and also the changes in (dual) volumes. Hence, for larger lattice sizes edge length changes affect more interactions and degrees of freedom compared to smaller lattices, which might lead to more disfavored proposals. This could be an explanation why for larger lattice sizes the tails in the length distribution might be shorter. Obviously, this is a peculiarity of the cuboid spin foam model and does not hold for general triangulations. It will be interesting to see whether a similar effect also occurs in such cases.

In summary, the sharply peaked length distributions suggest that in the plateau region, the spin foam on average looks like a regular lattice with average lattice spacing determined mainly by the scalar field mass $M$ and the parameter $\alpha$. Deviations around this regular shape do occur. In the next section we investigate the 2-point correlation function of the scalar field on this spin foam, and whether it is sensitive to the fluctuations around the regular lattice.

\subsection{Matter Observables}

The relevant matter observables of the coupled system are the 2-point correlation functions $\langle \phi_a \phi_b \rangle$ of the scalar field, where $a$ and $b$ denote vertices in the cubulation. Technically, this is not a pure matter observable: in addition to the correlation of the fields, we also measure their distance given by the specific spin foam configuration. This is crucial for the physical interpretation of the correlator, since the labelling of the vertices by itself has no physical significance and is merely a choice of coordinates. Of course, the combinatorics of the lattice plays a role, e.g. whether two vertices are neighbours and thus the associated fields interact directly. However, we do not track this information, as the distance appears to be the determining factor for the correlators.

In the following, we discuss the 2-point correlation functions in the ``plateau'' region for different lattice sizes. Investigating the correlators outside this region has little merit, for different reasons. Primarily, the results depend on the chosen upper and lower cut-offs and therefore cannot be trusted. Even if the cut-offs could be removed, sampling fields for infinitely small or large lattices is not possible or reasonable. Not to mention that the semi-classical approximation breaks down for spin foam configurations with tiny lengths.

\begin{figure}[h!]
    \centering
    \includegraphics[width=0.45 \textwidth]{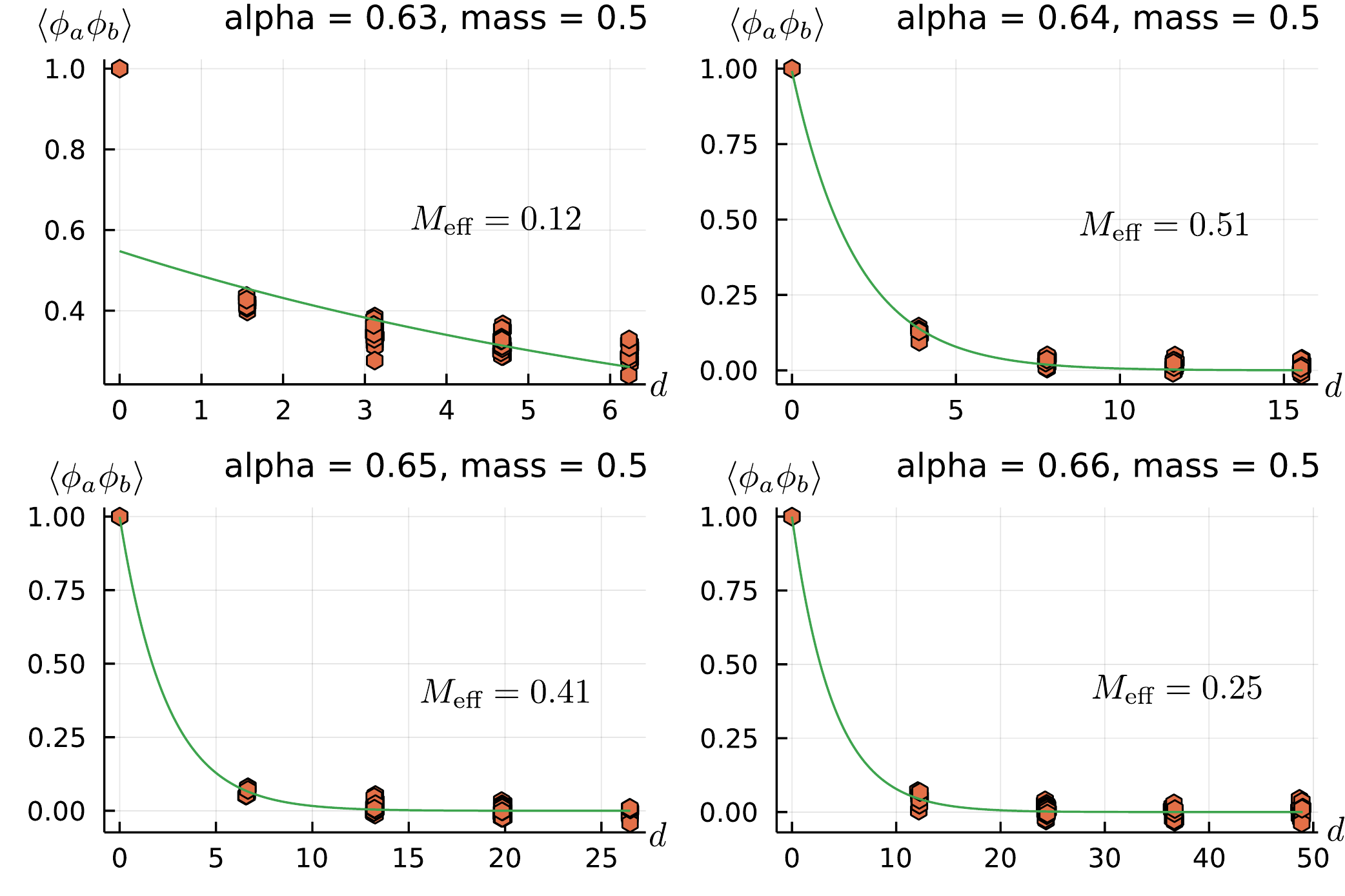} \; \;
    \includegraphics[width=0.45 \textwidth]{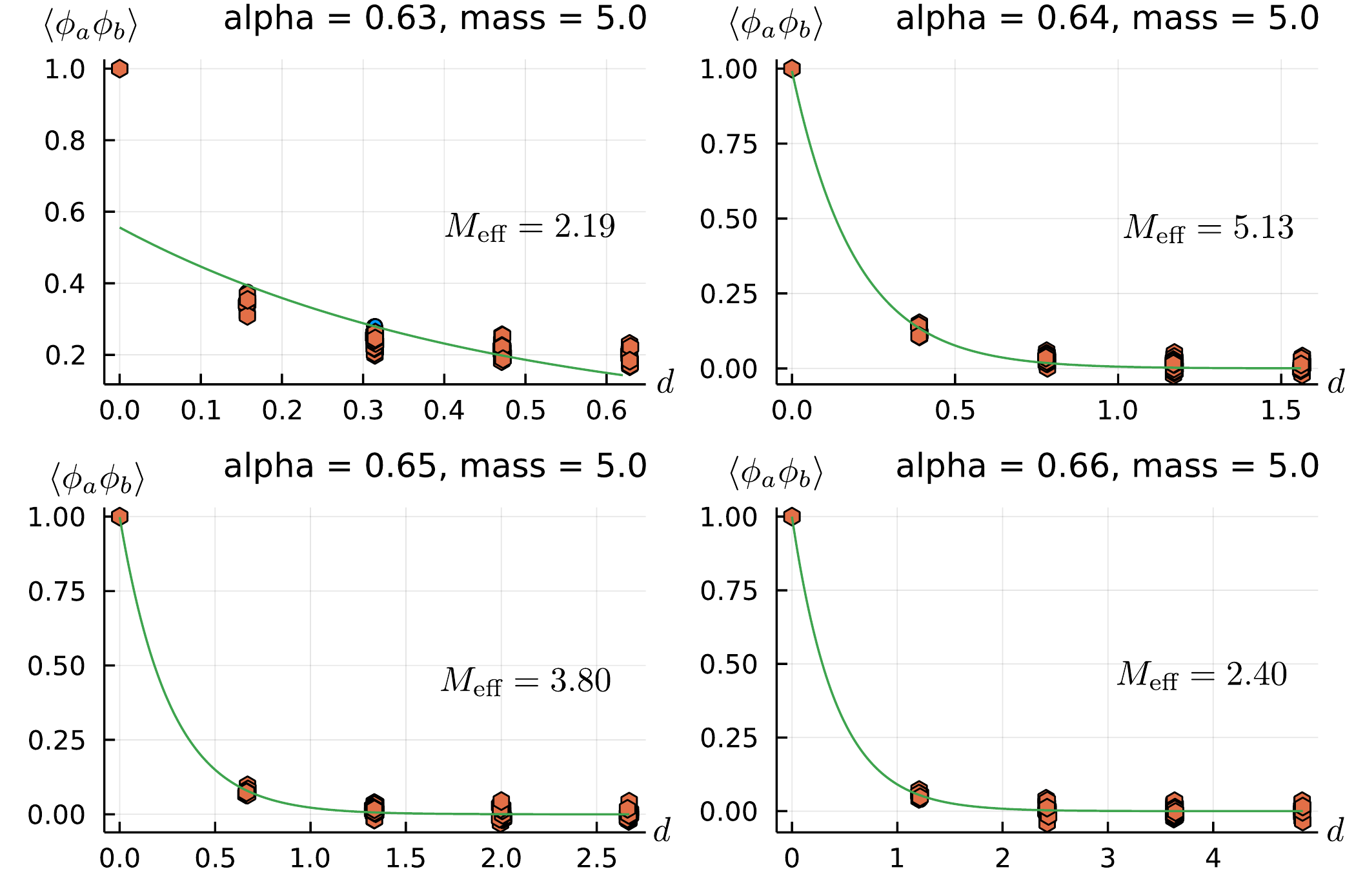}
    \caption{2-point correlation function $\langle \phi_a \phi_b \rangle$ plotted over the distance $d$ between the vertices $a$ and $b$ of lattices with $3$ independent edge length values ($N=81$).}
    \label{fig:size_3_corr}
\end{figure}

\begin{figure}
    \centering
    \includegraphics[width=0.45 \textwidth]{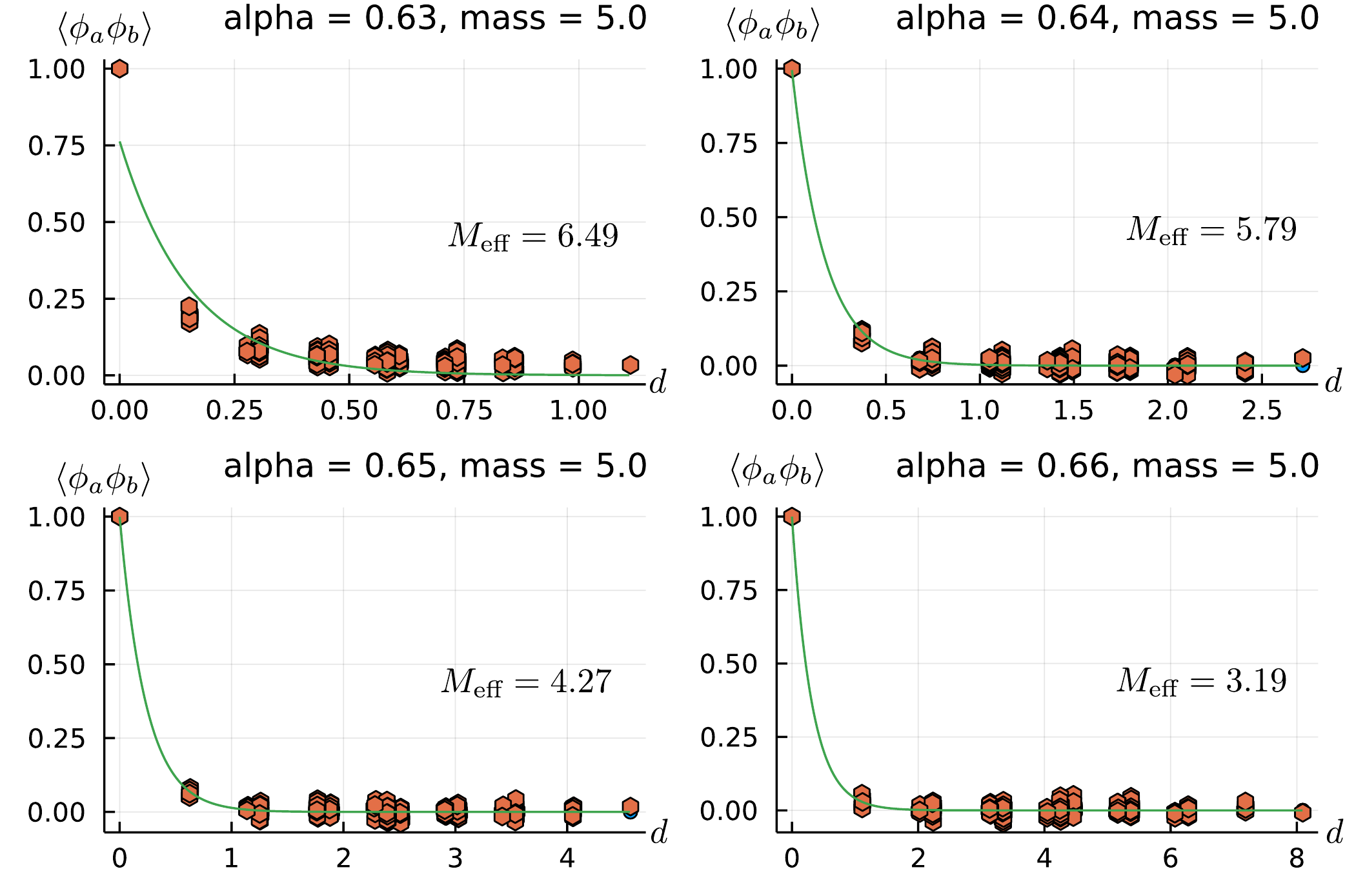} \; \;
    \includegraphics[width=0.45 \textwidth]{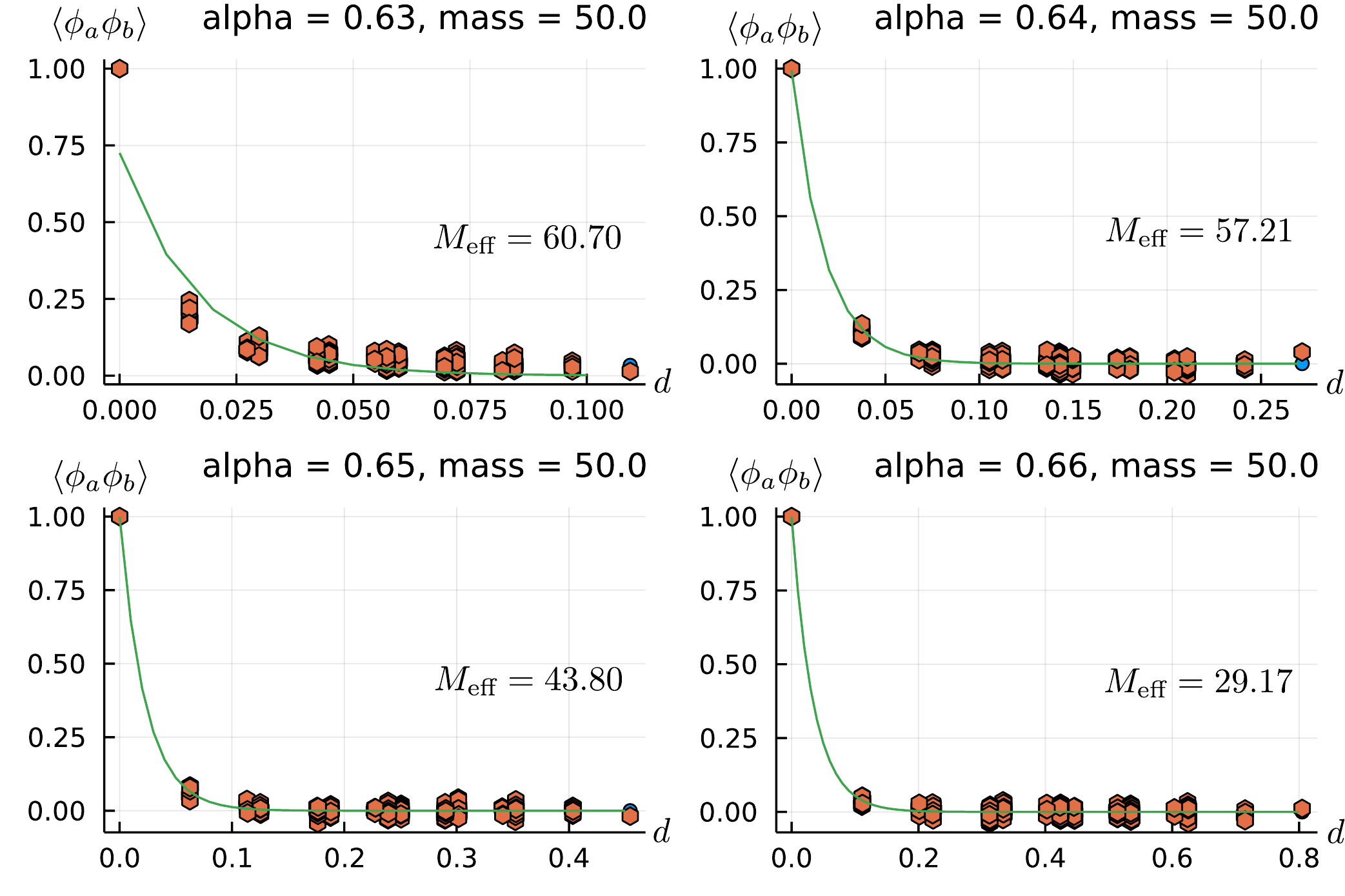}
    \caption{2-point correlation function $\langle \phi_a \phi_b \rangle$ plotted over the distance $d$ between the vertices $a$ and $b$ of lattices with $4$ independent edge length values ($N=256$).}
    \label{fig:size_4_corr}
\end{figure}

\begin{figure}
    \centering
    \includegraphics[width=0.45 \textwidth]{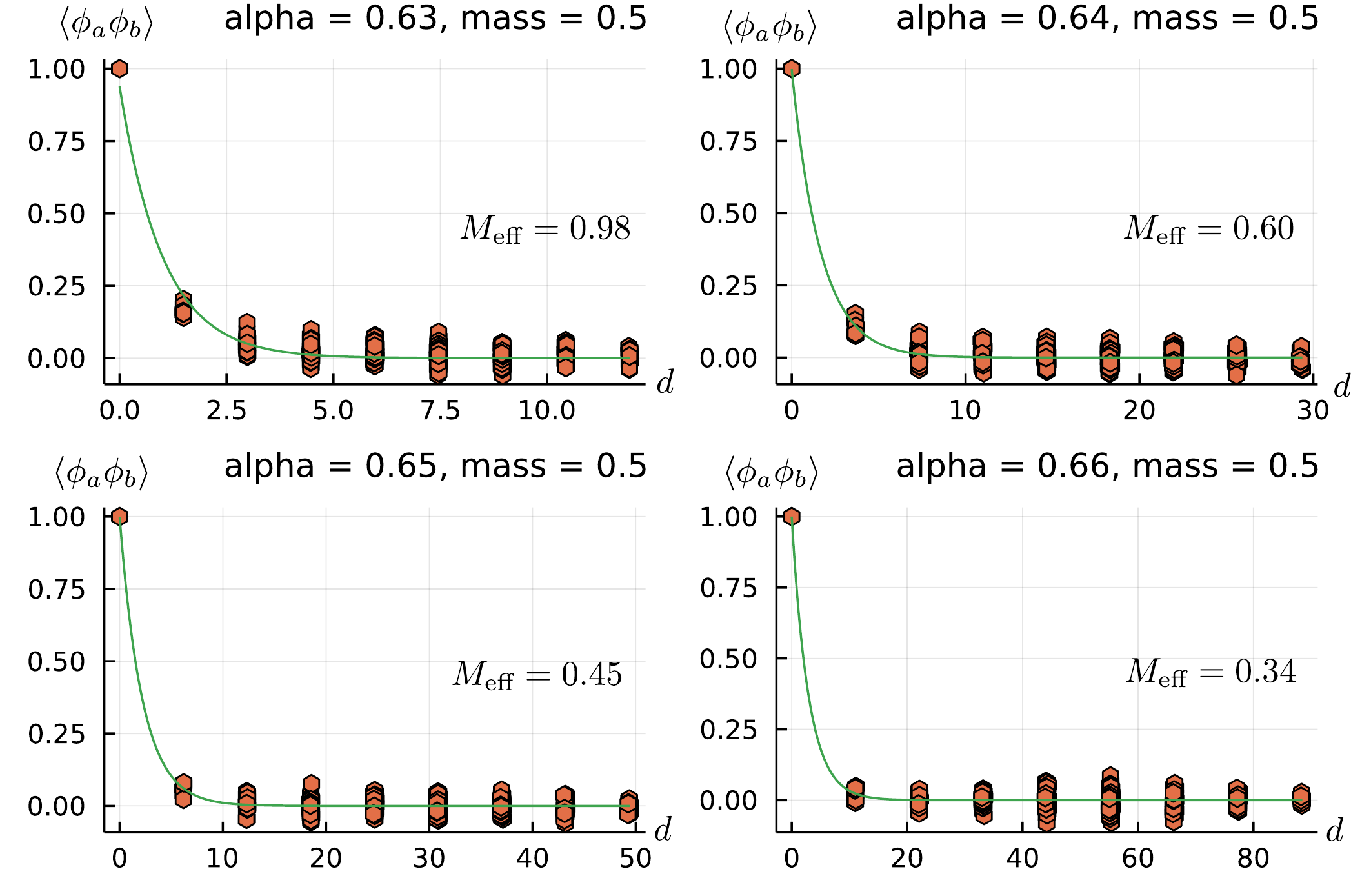}
    \caption{2-point correlation function $\langle \phi_a \phi_b \rangle$ plotted over the distance $d$ between the vertices $a$ and $b$ of lattices with $5$ independent edge length values ($N=625$).}
    \label{fig:size_5_corr}
\end{figure}

In fig. \ref{fig:size_3_corr}, \ref{fig:size_4_corr} and \ref{fig:size_5_corr} we plot the samples of the 2-point correlators $\langle \phi_a \phi_b \rangle$ for all pairs of vertices $a$, $b$ over the distance that vertices $a$ and $b$ are apart (in the respective sample), while also dropping the labels. That way we consider scalar field correlations as a function of distance. We define the length between two vertices as the length of the shortest path between them, i.e. we sum up the edge lengths along this path. Note that there are several paths connecting two vertices due to periodic boundary conditions. 
Thus we consider field correlations as a function of distance between the fields, similar to lattice field theory, yet these distances are dynamical and determined by the gravitational field encoded in the spin foam. Hence, we are measuring a relational observable \cite{Rovelli:2001bz,Dittrich:2005kc,Tambornino:2011vg}.

Following our discussion in the previous section, let us first discuss the distances of vertices in the correlation plots. For lattice sizes $N=81$ and $N=625$, see figs. \ref{fig:size_3_corr} and \ref{fig:size_5_corr} respectively, we see that the correlations are are measured at almost equidistant distances with barely any spread, which is difficult to resolve at the plotted distances. This observation is in agreement with the distribution of lengths, and shows that effectively the spin foam behaves like a regular lattice. Comparatively, for lattice size $N=256$ in fig. \ref{fig:size_4_corr} the correlations appear in ``bunches'', but in a distinct pattern: closest and furthest neighbours appear only in a single line, the intermediate one in roughly three lines. Since this consistently occurs in all simulations for lattices with $4$ independent edge lengths but not for lattices with $3$ and $5$ independent edge lengths, and we find qualitatively similar length distributions for all three lattice sizes, this is most likely a peculiarity of lattices with an even number of independent edge lengths. Due to the cuboid symmetry and periodic boundary conditions, there exist 16 different paths (of different lengths) between two vertices, essentially the two directions per dimension. On odd lattices, there always exists a combinatorially shortest path, which immediately translates to (almost) regular lattices. On even lattices, however, certain pairs of vertices can have multiple paths of the same combinatorial distance between them. In these cases, the fluctuations around regular lattices become crucial and thus might allow for more deviations of the distance. Hence, the pairs of vertices which have an intermediate distance between them show the largest spread in their distance, while furthest away pairs are sharply peaked. Unfortunately, we cannot confirm this for a larger even lattice, e.g. lattices with $6$ independent edge lengths per dimension, as the numerical costs grow drastically. These results furthermore support our hypothesis that on average, the spin foam appears as a regular lattice with fluctuations around this regular shape.

For the correlations themselves, we make two main observations. Firstly, the correlations fall off quickly and are typically below $1/4$ already for the smallest non-vanishing distance. This is mainly due to magnitude of this distance, which is dynamical. As discussed above, while smaller edge lengths are permitted by the spin foam model, they are highly improbable and thus it is extremely  difficult to measure the correlations at small distances. On the other hand, because these configurations are so unlikely, we can assume that the correlations of the scalar field for smaller non-vanishing distances are highly suppressed, but more research is necessary to explore this question further. The second observation is that there is significant spread in the correlations also at large lengths. At such distances, the system actually is difficult to sample as only extremely small values of the scalar field are viable. However, due to these large deviations on the correlations and finite sample size, the correlations might on average appear to be larger than expected for such distances in ordinary lattice field theory. This might affect the estimates of the mass from the exponential decline of the correlators.

Qualitatively, the results look strikingly similar for different masses $M$, but careful attention is necessary: For all sets of parameters, we qualitatively observe that correlations exponentially decay as the distance between vertices grows. This observation is true for any lattice size, mass of the scalar field $M$ and value of $\alpha$ (in the plateau region). However, $\alpha$ and $M$ play starkly different roles. $\alpha$ is a parameter that only appears in the spin foam amplitude and it is directly related to the lengths: larger $\alpha$ implies a larger probability of larger lengths, meaning that the distance between vertices $a$ and $b$ increases, which in turn leads to lower correlations. Conversely, a larger mass implies a shorter correlation length, however at first sight the correlations do appear to decay similarly to the cases with lower mass. This is due to the change in length; a larger mass favours smaller lengths, such that the vertices are on average closer together. Thus, the measured correlation length is smaller, due to larger $M$, yet the fields are similarly correlated since the vertices are effectively closer. Hence, the scalar field and in particular the value of its mass has a strong impact on the geometry of space-time. On the other hand, the 2-point correlation function on average behaves like the 2-point correlation function of a scalar field theory of the same mass on a regular lattice, whose lattice spacing agrees with the average lattice spacing of the spin foam. This also holds for any lattice size, where essentially a larger lattice size adds pairs of vertices to the system that are further apart. To quantify our impression that the 2-point correlator of the scalar field is barely affected by the dynamical spin foam, we compute the correlation length, i.e. the inverse of the ``effective'' mass, from the exponential decay of the correlator.

In scalar lattice field theory, the 2-point correlation function exponentially decays with the distance between the vertices, where the decay rate is called the correlation length. For a free, massive scalar field, this correlation length is equal to the inverse mass of the scalar field; a larger mass implies a shorter correlation length. In our setup, we can thus disentangle changes of the correlations and lengths by fitting the exponential decay of the correlation function and infer an effective mass of the scalar field coupled to the spin foam. Note that this is also a coarse, non-local observable derived from the entire system.

\begin{figure}
    \centering
    \includegraphics[width=0.45\textwidth]{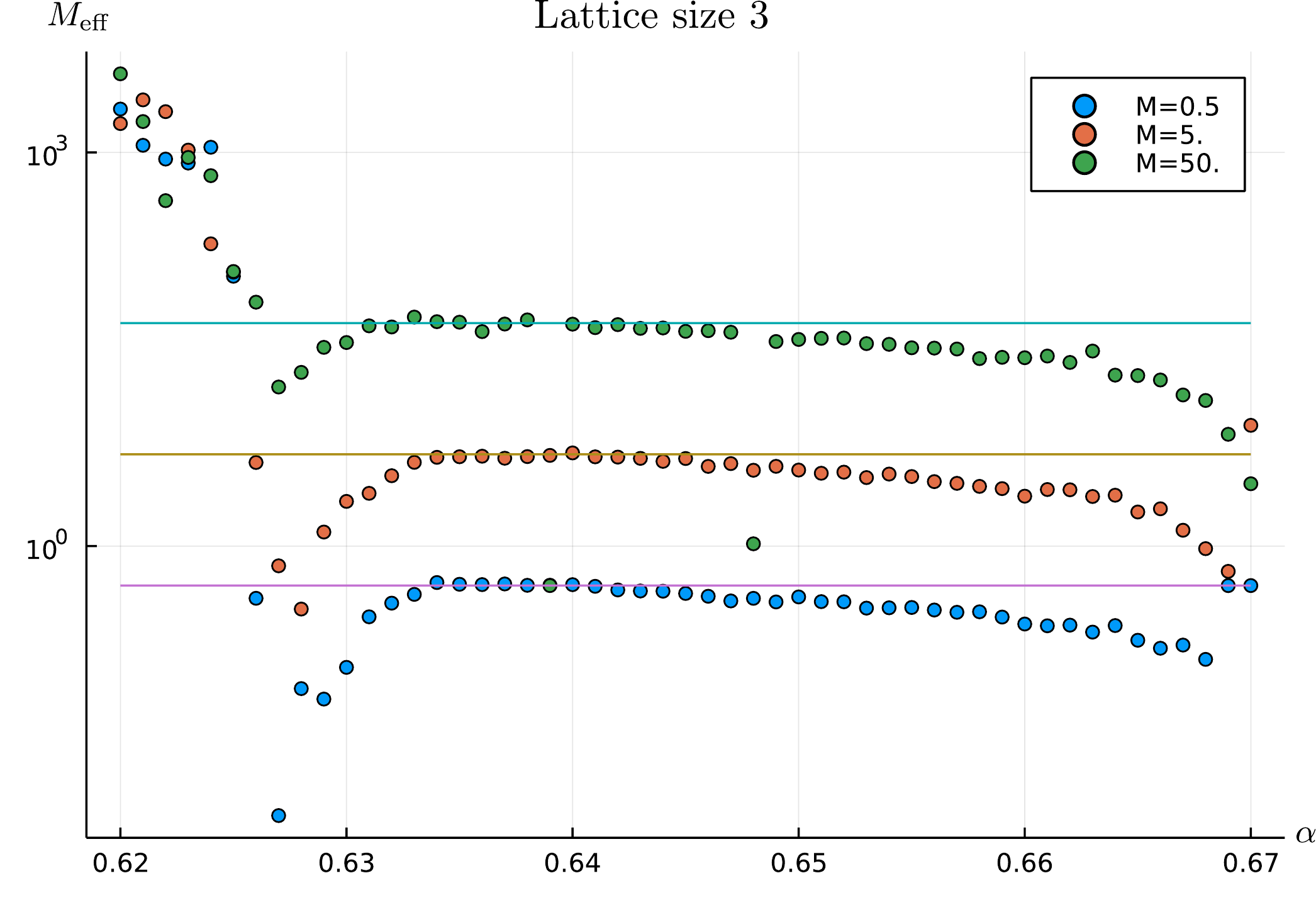} \; \;
    \includegraphics[width=0.45\textwidth]{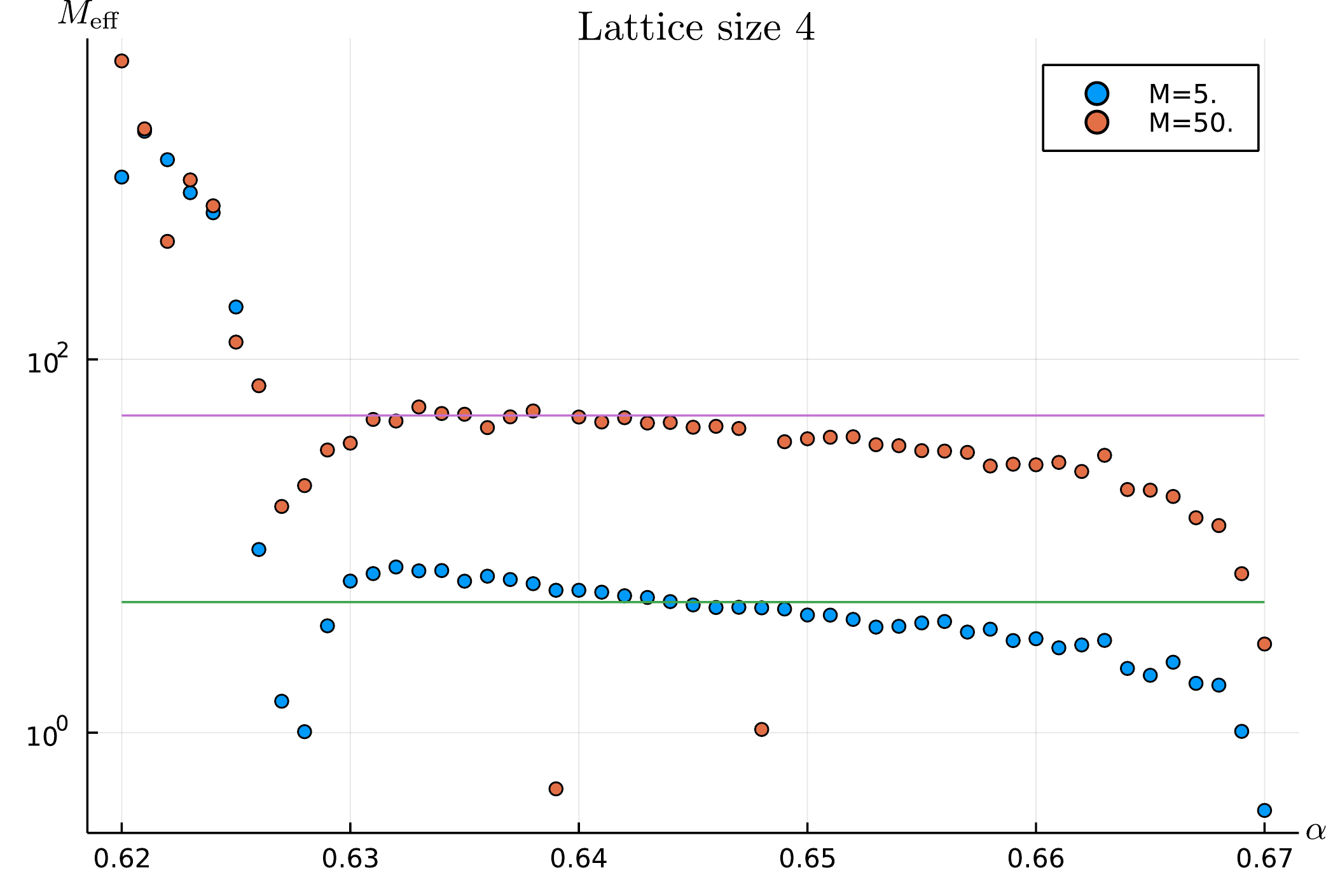}
    \caption{Plots of mass estimated from exponential decline of scalar field correlations in logarithmic scale. Horizontal lines show the mass parameter of the model for comparison. \textit{Left:} lattice size 3. \textit{Right:} lattice size 4.}
    \label{fig:eff_mass_plots}
\end{figure}

\begin{figure}
    \centering
    \includegraphics[width=0.45\textwidth]{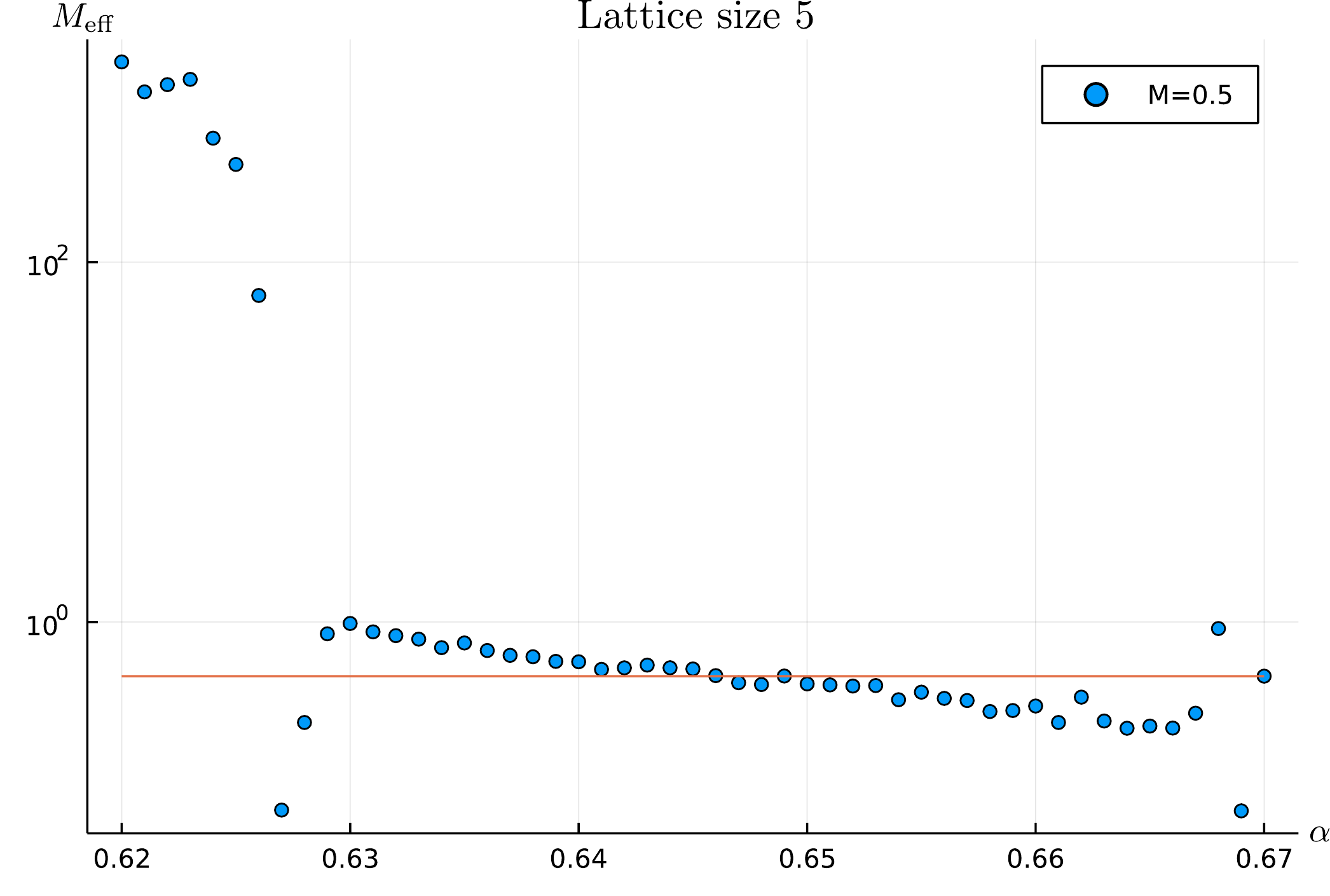}
    \caption{Plot of mass estimated from exponential decline of scalar field correlations in logarithmic scale for lattice size 5. Horizontal lines show the mass parameter of the model for comparison.}
    \label{fig:eff_mass_plot_5}
\end{figure}

In figs. \ref{fig:eff_mass_plots} and \ref{fig:eff_mass_plot_5} we plot the measured effective mass for a fixed mass $M$ of the scalar field as a function of the parameter $\alpha$ and different lattice sizes. First of all, we see that in the plateau region the effective mass is at a similar order of magnitude as the field mass $M$, yet it changes with $\alpha$. For $\alpha$ around $0.64$, we see a good agreement between the fitted and actual mass. However, the effective mass decreases as we increase $\alpha$, to around half the value of the field mass close to the end of the plateau. Outside the plateau region, the mass estimates from the fits cannot be trusted, because either the lattice is close to the lower or upper cut-off respectively. However, the decline of the mass in the plateau region towards larger $\alpha$ remains to be understood.

The decline of the effective / fitted mass is universal for different lattice sizes and masses. Since this decrease is tied to $\alpha$, the cause might be related to changes in the spin foam, e.g. the length distribution. For larger $\alpha$, we observe a broader spread in the length distribution with a longer tail towards larger lengths. However, it is not clear how this would translate into correlations that die off later and thus result into a smaller effective mass. Moreover, for larger lattice sizes this broadening of the length distribution is less pronounced, yet the decline in mass is comparable. Instead a different explanation appears more likely. 


As discussed above, we observe a significant spread of correlations also for large distances, even though the correlations are exponentially suppressed. This is due to the fact that scalar lattice field theory is difficult to sample in those regions. Due to this systematic issue of Monte Carlo simulations, the correlations might appear on average larger than they should be, from which we obtain a larger correlation length / smaller effective mass from the fit to the measured 2-point correlation function. To our best knowledge, this is the most plausible explanation for the observed effect, which makes it unlikely that this difference between mass parameter $M$ and effective mass is a physical effect. Certainly, this question for the cuboids should be investigated in more detail. Furthermore, it will be interesting to see whether this result still holds for more general spin foams, which allow for curved geometries.

To summarize, the observables of the scalar field coupled to cuboid spin foams are remarkably similar to studying a scalar field on a regular lattice, giving further evidence to our hypothesis that our model can be effectively regarded as a scalar field on a regular lattice with dynamical lattice spacing. Changing the mass of the scalar field mostly affects the spin foam, while the correlation function remains similar to one on a regular lattice. Estimating the scalar field mass from the correlation function does reveal a decline with $\alpha$, yet it seems unlikely that this is a physical effect and rather rooted in the difficulty of sampling the scalar field for large lattice lengths.

\section{Summary and discussion} \label{sec:discussion}

In this article we have defined the coupling of free, massive scalar field theory to a 4d (semi-classical) spin foam model and for the first time computed expectation values of this coupled matter gravity system in the restricted case of cuboid spin foams \cite{Bahr:2015gxa}. The idea of the coupling mechanism is to consider a spin foam configuration as an irregular lattice on which we define lattice field theory via discrete exterior calculus \cite{desbrun_marsden,Calcagni:2012cv,Thurigen:2015paa}. Then, we sum over spin foam configurations and thus superimpose lattice field theories weighted by spin foam amplitudes, akin to the ideas developed in \cite{oriti-pfeiffer}. Our ansatz is not restricted to cuboids and can be straightforwardly adapted to other semi-classical approaches like frusta spin foams \cite{Bahr:2017eyi} or effective spin foams \cite{Asante:2020iwm}\footnote{For effective spin foams special attention must be paid to configurations, where two 4-simplices are glued along a tetrahedron, yet the shapes of this tetrahedron seen from the two 4-simplices do not match.}. The particular advantage of cuboid spin foams is that its amplitudes are non-oscillatory, even in the quantum regime \cite{Allen:2022unb}, such that we can use Markov Chain Monte Carlo techniques to compute observables of the matter spin foam system.

In this paper, we have studied two types of observables, geometric ones, like the total volume and edge lengths distribution, and matter ones, like the 2-point correlation function. From these observables, we identify two counteracting mechanisms of the coupled matter gravity system: the scalar field leads to a polynomial suppression of large lengths, derived from integrating out the scalar fields, which is more pronounced for larger mass. On the other hand, the parameter $\alpha$ in the spin foam face amplitude favours large lengths as it is increased. From the geometric observables, we then identify three regimes: for small $\alpha$, the lengths of the system are as small as possible, while for large $\alpha$ the lengths are as large as possible. These two regimes are connected by an extended region in parameter space at which the total volume and the edge lengths are finite, and the system on average / effectively looks like a regular lattice. The average lattice size is a function of the scalar field mass $M$ and $\alpha$ and appears to be independent of the lattice size. Note that this lattice size is \textit{dynamical} and not related to the fundamental discreteness of area variables in spin foams; due to the semi-classical regime we have assumed the lengths variables to be continuous. Additionally, we studied the 2-point correlation function of the scalar field, which effectively behaves as if the scalar field is defined on a regular lattice. While our estimates show some deviations in the effective mass / correlation lengths, we attribute those to the difficulty of measuring correlations for large lattice distances. Interestingly, changing the mass of the scalar field mostly appears to affect the spin foam, whose average edge lengths change inversely proportional to the change of mass.

Our results have several interesting implications for spin foam quantum gravity. Primarily, the effective \textit{emergence} of scalar lattice field theory defined on a dynamical regular lattice is first evidence that spin foam quantum gravity coupled to matter might possess a regime that effectively describes quantum field theory defined on a fixed background space-time. Additionally, the effective lattice spacing is also dynamical and given as a function of the parameters of the theory, in particular the scalar field mass. This could hint towards a new mechanism as to how new intermediate scales can emerge in spin foams / loop quantum gravity coupled to matter. It furthermore raises the questions as to how other types of matter will influence the spin foam, and whether there are limits to how many matter fields we can couple and still recover a quantum field theory from our coupled system. This is indeed a question raised in many approaches to quantum gravity, e.g. asymptotic safety \cite{Eichhorn:2022jqj}, in particular whether the standard model of particle physics is compatible and whether there is a mechanism that limits the amount of matter fields one can couple to quantum gravity.

Additionally, in this work we gain first insights into how to extract physical observables from matter-spin foam systems. In absence of a background space-time, no overall scale like a lattice spacing is a priori available to define a correlation lengths. One can resort to using combinatorial information or simply the labels of vertices, but this is a coordinate distance instead of a physical one. Instead we measure both the distance (here path distance) between two vertices as well as the correlation of the scalar field located at these vertices and drop their labels. That way we store physical, diffeomorphism invariant information, namely correlations in relation to the geodesic distance between two points, in the spirit of relational observables \cite{Rovelli:2001bz,Dittrich:2005kc,Tambornino:2011vg}. Note however, that in our Monte Carlo approach, we cannot probe the correlations at arbitrary distances, only at \textit{probable} distances. In the review on causal dynamical triangulation \cite{Ambjorn:2012jv} a scalar field 2-point correlation function is discussed that defines the correlator at arbitrary geodesic distances.

At this stage it is not clear how to define a continuum / refinement limit \cite{time-evo,Steinhaus:2020lgb} of the coupled spin foam matter system. From the observables we studied in this work, none appears to be an order parameter indicating a second order phase transition. While the variance of the total volume shows a divergent behavior at the edges of the plateau region, these divergences do not grow for larger lattice sizes. Moreover, it would be necessary to evaluate in more detail whether the variance is dependent on the chosen cut-off. Beyond the conclusions we can draw from our current work, renormalization / coarse graining of the matter gravity system, similar to \cite{Steinhaus:2015kxa}, is indispensable to address this question and to evaluate what influence various ambiguities might have on the dynamics, e.g. the definition of the scalar field theory. In our concrete case, we could attempt to define a renormalization group flow where the system defined on a finer lattice effectively looks like the refinement of a coarser system, while keeping the 2-point correlation function of the scalar field fixed. Essentially this amounts to matching the average total volume by tuning the parameter $\alpha$ and keeping the mass $M$ fixed. In the plateau region, where the 2-point correlation function and correlation length are well understood, this strategy of identifying observables will likely only work for a few coarse graining steps. On the other hand, on the edges of the plateau region, where the lengths start to fluctuate, it might be possible to relate the average geometric observables across different lattice sizes by tuning $\alpha$. Further work is necessary to explore this idea, in particular how the scalar field correlator behaves for such parameters on larger lattices.

While the results of this article a encouraging and intriguing, it is clear that many assumptions are made that must be lifted in order to check whether they also hold in physically more relevant situations. On the side of the spin foam this requires us to go beyond quantum cuboids and eventually beyond semi-classical models. A suitable first step could be to use effective spin foams \cite{Asante:2020iwm} defined on a triangulation, which is numerically more accessible than the full quantum theory and allows for vastly more configurations with non-vanishing curvature and non-metricity. However, due to the oscillatory nature of the amplitude, Monte Carlo methods will no longer be applicable and other numerical methods must be explored. A first attempt might be to consider a scalar field coupled to effective spin foams expanded around a flat triangulation of hypercubes similar to the work \cite{Dittrich:2022yoo}. Extending the matter coupling to the quantum regime of spin foams is another interesting challenge that must be addressed, as well as coupling other types of matter and studying their dynamics, e.g. gauge field \cite{oriti-pfeiffer} and fermions \cite{Bianchi:2010bn,Han:2011as}.

\section*{Acknowledgements}
The authors would like to thank L Glaser for insightful discussions on Markov Chain Monte Carlo techniques and Johannes Th\"urigen for valuable insights into lattice Laplacians on irregular lattices. Furthermore, the authors are grateful for comments and insights on matter couplings in quantum gravity by Astrid Eichhorn and Viqar Husain.
S.St. is funded by the Deutsche Forschungsgemeinschaft (DFG, German Research Foundation) - Projektnummer/project-number 422809950. During the course of this work, M.Ali was supported by the Bryce DeWitt Fellowship as part of the Discretuum to Continuum Initiative and by the Natural Sciences and Engineering Research Council of Canada (NSERC) through a Post-Doctoral Fellowship.
This research was in part supported by Perimeter Institute for Theoretical Physics.
Research at Perimeter Institute is supported in part by the Government of Canada through the Department of
Innovation, Science and Economic Development Canada and by the Province of Ontario through the Ministry of
Colleges and Universities.

\bibliographystyle{utphys}
\bibliography{bibliography}

\end{document}